# Chapter 10: The Application of Artificial Intelligence in Software Engineering – A Review Challenging Conventional Wisdom

**Feras A. Batarseh**

College of Science

George Mason University

Fairfax, VA, USA

**Rasika Mohod**

College of Engineering

George Mason University

Fairfax, VA, USA

**Abhinav Kumar**

College of Engineering

George Mason University

Fairfax, VA, USA

**Justin Bui**

Department of Electrical Engineering and Computer Sciences

University of California, Berkeley

Berkeley, CA, USA

*"The less there is to justify a traditional custom, the harder it is to get rid of it" – Mark Twain*

**Abstract** – the field of Artificial Intelligence (AI) is witnessing a recent upsurge in research, tools development, and deployment of applications. Multiple software companies are shifting their focus to developing intelligent systems; and many others are deploying AI paradigms to their existing processes. In parallel, the academic research community is injecting AI paradigms to provide solutions to traditional engineering problems. Similarly, AI has evidently been proved useful to Software Engineering (SE). When one observes the SE phases (Requirements, Design, Development, Testing, Release and Maintenance), it becomes clear that multiple AI paradigms (such as: Neural Networks, Machine Learning, Knowledge-Based Systems, Natural Language Processing) could be applied to improve the process and eliminate many of the major challenges that the SE field has been facing. This survey chapter is a review of the most commonplace methods of AI applied to SE. The review covers methods between years 1975-2017, for the requirements phase, 46 major AI-driven methods are found, 19 for design, 15 for development, 68 for testing, 15 for release and maintenance. Furthermore, the purpose of this chapter is *three-fold*; firstly, to answer the following questions: is there sufficient intelligence in the SE lifecycle? What does applying AI to SE entail? Secondly, to measure, formulize, and evaluate the overlap of SE phases and AI disciplines. Lastly, this chapter aims to provide serious questions to challenging the current *conventional wisdom* (i.e. status quo) of the state-of-the-art, craft a call for action, and to re-define the path forward.

**Keywords:** Lifecycle Phase; Artificial Intelligence Paradigms; Requirements; Testing; Design.

# 1. Introduction and Motivation

In the 1880s, Charles Babbage built a machine that was capable of performing an assortment of mathematical calculations. The goal of the machine was to execute math correctly. Babbage's purpose was to get rid of the inherent errors that occur when humans do calculations by hand [1]. Many refer to that program, as the first software program. Software, since its earliest stages, aimed to help humans automate processes that require a certain level of 'intelligence'. The process of building software (i.e. Software Engineering), remains primarily a human activity. Within computer science, the field of Software Engineering (SE) is the dominating industrial field [2]. Nonetheless, there has been ample amounts of research in every aspect of the SE process. As the field advances, many important questions were still not answered with consensus, such as: How to estimate time and cost of a project? When to stop testing the system? How to locate and eradicate errors? How to incrementally develop and re-design? How to refactor the code? How to turn requirements into features? And many other important questions.

Similar to other engineering disciplines, building a software follows a well-defined process, this process is referred to as a lifecycle. Many different lifecycles have been introduced in literature, and many have been used in industry. Commonplace lifecycle models include: waterfall, agile, spiral, rapid, and incremental. Each phase within the lifecycle has its own challenges, drawbacks, and best practices; each phase therefore, became its own research field that has communities of researchers trying to improve it, and multiple conferences and journals trying to advance it. In parallel, another area of research that is meant to represent intelligence in a machine, is Artificial Intelligence (AI). Many researchers claimed that these two areas of research (SE and AI) haven't interacted or overlapped sufficiently [1][2][3][4]. Software's aim is to model the real world, to represent a certain human activity, or to automate an existing manual process. Similarly, AI - within its many subfields - aims to solve problems, represent intelligence, engineer knowledge, recognize patterns, learn from experience, and eventually answer the famous questions posed by Alan Turing: 'Can Machines Think?', think *intelligently* that is [1]. Intelligence, based on a recent definition from the Artificial General Intelligence proceedings [3][4], is the ability to "perform complex tasks within a complex environment". In 2008, Rech et al claimed that: "the disciplines of AI and SE have many commonalities. Both deal with modeling real world objects from the real world like business process, expert knowledge, or process models" [3]. Hence, there is no doubt that the disciplines of SE and AI have a lot to share.

To evaluate the success of AI when applied to SE, an updated and extensive review of the state-of-the-art is overdue (therefore, this chapter). There seems to be a general consensus that AI is suitable for solving SE problems [5][6][7]. There are many methods that apply AI to SE, however, most of the time, the application is theoretical, small, mathematical, or applied to a formal problem within SE. No paper was found that addressed the non-functional aspects of SE for instance, the human face of software engineering, communication issues between teams, or the 'real world' reasons why SE projects fail. For example, a method that intelligently classifies or clusters test cases in testing might be helpful in some cases, but it still wouldn't answer the *big* and ongoing questions of testing (what is a valid software? When does the SE team claim an errorless system?). SE researchers were able to address traditional *formal* issues, however, is AI able to solve bigger problems in SE? That is still unclear. This chapter will address this question. The purpose of the argument presented is not to search for a silver bullet, rather, it is to challenge the many papers that claimed that the solution of many traditional SE challenges lies in AI. Most importantly however, very few methods AI-driven have been used in industry. The real evaluation of such success should be reflected in the number of success stories, and implementations of these methods in industry.

There has been a number of reviews that covered the same topic [4][3], however, none of the ones found are as extensive, or as detailed as this manuscript. Additionally, in this chapter, the methods are broken down by SE lifecycle phase. Five phases are used: 1- Requirements Engineering 2- Design 3- Development 4- Testing 5- Release and Maintenance. These phases are deemed to be the most 'commonplace' ones [4], furthermore, while searching for papers, these phases proved relevant to how the research community is divided. Most researchers –when writing about a certain phase– refer to one of these phases. Moreover, some papers per phase were found but not included in this review (they were deemed less important based on citations and relevance to scope). Some papers were more relevant and had a clear application of AI to a SE phase, while others were a bit vague onto which phase they belong to, or the AI paradigm that they use. Nonetheless, and to provide a complete review -of all papers found- in this area of study, a comprehensive **list** of papers is provided, including the *supposed* phase that it belongs to, and the AI paradigm that it follows. That is discussed further within sections 2 and 3.

This chapter is structured as follows: the following five sub-sections review AI for each of the five SE lifecycle phase (which constitutes section 2); section 3 provides a summary of the review (including the complete list of all AI-driven SE methods); and section 4 concludes the manuscript, includes insights, and ideas for the path forward.

## 2. Applying AI to SE Lifecycle Phases

This section has five sub-sections, each one is dedicated to a SE phase and its AI methods. Within each phase, multiple AI-inspired methods are reviewed. The review within each section is neutral and unbiased as it merely provides the claims that were made by the original referenced authors. Contrary to that, the subsequent sections of this survey (3 and 4) provide arguments, insights and further discussions.

## 2.1 Requirements Engineering and Planning

Requirements analysis and planning is the first stage in the software engineering process and it forms the building blocks of a software system. Software requirements describe the outlook of a software application by specifying the software's main objectives and goals. Due to its importance, many researchers tried to impose AI into this generally first phase of the lifecycle. ([5][6][7][8][9][10][11][12][13][14][15][16][17][18][19][20][21][22][23][24][25][26][27][28][29][30][31][32][33]).

The Institute of Electrical and Electronics Engineers (IEEE) defines **a requirement** as a condition or capability needed by a user to solve a problem or achieve an objective, which must be met by a system to satisfy a contract, standard, specification, or other formally imposed document [34]. Requirements thus define the framework for a software development process, by specifying what the system must do, how it must behave, the properties it must exhibit, the qualities it must possess, and the constraints that the system must satisfy. Requirements Engineering (RE) is the engineering discipline of establishing user requirements and specifying software systems. RE emphasizes the use of systematic and repeatable techniques that ensure the completeness, consistency, and relevance of the system requirements. RE is considered as one of the most critical phases of a software development life cycle as the unclear, ambiguous, and low-quality requirements specification can lead to the failure of a product or deployment of a completely undesired product and even raise development costs. Therefore, and due to the increasing size and complexity of software systems, there is a growing demand for AI approaches that can help to improve the quality of the RE processes. The research work concerning the need for computer-based tools which help human designers formulate formal and process-oriented requirements specifications date as back as 1978 [35]; Balzer et al. determined some attributes of a suitable process-oriented specification language, and examined why specifications would still be difficult to write in such a language in the absence of formulation tools. The researchers argued that the key to overcoming these difficulties was the careful introduction of informality based on partial, rather than complete, descriptions and the use of a computer-based tool which utilizes **context** extensively to complete these descriptions during the process of constructing a well-formed specification. The objects were perceived entirely in terms of their relationships with each other and set of primitive operations which allow relationships to be built or destroyed. This allowed incremental elaboration of objects, relationships, and operations. This approach allowed objects and operations to be modelled almost exactly as a user desires. However, constructing operational specification still proved to be difficult and error prone because the specification language was still a formal language. Specification languages only provided partial descriptions because the context provided the rest of the necessary information. The authors further proposed the context mechanisms, which were more complex, and produced more diffused contexts. There had also been a concern over the difficulty and reliability of creating and maintaining specifications over the program's lifecycle. The authors proposed a tool that assisted in converting informal specifications into formal ones [35]. Advantages of informal specifications included their precision, helped focus development, reduced training for users, and assisted with maintainability. The authors proposed an idea where any changes needed to be made to specifications are made to the informal specifications because they would be much easier to deal with. The authors tested this new idea using a prototype system (called SAFE). The system was given an input for some tasks, and then the prototype produced a program. From the examples presented in the paper, the prototype performed successfully and errors found could be reduced with a little bit of user's input. Many other partly automatic tools have been developed to assist SE in the requirement analysis phase (such as: textual descriptions and Unified Modeling Language (UML) diagrams). To understand the dynamics of a system, a developer needs to analyze the use case descriptions and identify the actors before modeling the system. The common technique is to use grammar in the elicited text as the basis for identifying useful information. However, there is a scalability issue that is due to the *unstructured* nature of Natural Language (NL). A semi-automatic approach to extracting required information using **Natural Language Processing** (NLP) reduces the time spent on requirements analysis. From a computational perspective, the text can be a set of letters, words and sentences arranged in different patterns. Each pattern of words or sentences gives a specific meaning. Vemuri et al. [36] presented a probabilistic technique that identifies actors and use cases that can be used to approach this problem. Their research aims to explore **Machine Learning** techniques to process these patterns in the text (through NLP as well). A supervised learning algorithm is used to classify extracted features as actors and use cases. Additionally, input text is provided to a natural language processor to obtain word features. An extraction algorithm predicts use cases in a subject-verb-object combination. These results are used to draw use case diagrams. The process flow of the proposed approach is shown in Figure 1.

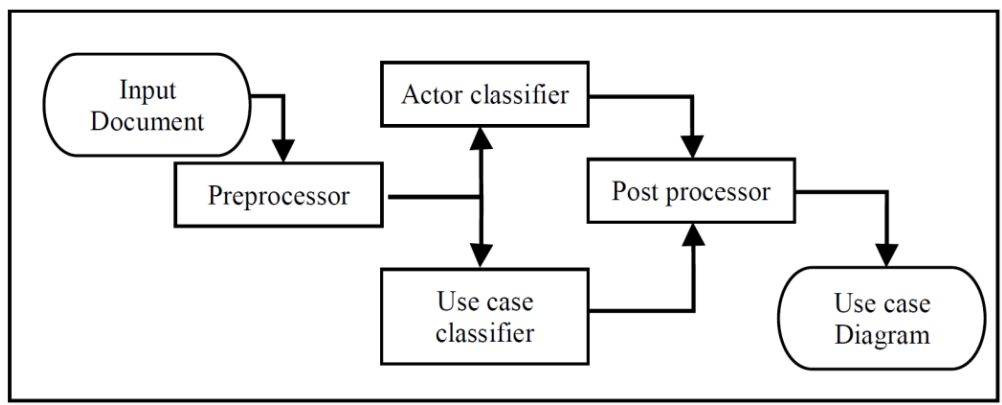

*Figure 1: Intelligently Processing Requirements Text [36]*

Experimental work in the paper [36] successfully attempted to extract actors and use cases using a probabilistic classification model along with minimal assistance from a **Rule-Based** approach.

The use cases obtained were crisp and consistent, irrespective of the size of requirements text. R-Tool, NL-OOPS, CM-BUILDER are few other NLP based computer aided software engineering tools [37]. Such tools produce class diagrams from the user requirement document (although it still requires user intervention). Michl et al. [37] proposed the NL-OOPS (Natural Language – Object-Oriented Production System) project to develop a tool supporting object oriented analysis. Requirements documents are analyzed with LOLITA (Large-scale Object-based Linguistic Interactor, Translator and Analyzer), a large-scale NL processing system. Both, the knowledge in the documents and ones that are already stored in the **Knowledge-Base** of LOLITA are then proposed to be used for producing requirements models. The approach was based on the consideration that requirements are often written in unrestricted NL and in many cases, it is impossible to impose customer restrictions on the language used. The object-oriented modelling module implemented an algorithm that filters entity and event nodes in its knowledge base to identify classes and associations. In parallel, Ninaus et al. [38] proposed a method to reducing the risk of low-quality requirements through improving the support of stakeholders in the development of RE models. The authors introduced the INTELLIREQ environment. This environment is based on different recommendation approaches that support stakeholders in requirements-related activities such as definition, quality assurance, reuse, and release planning. There are four basic types of recommendation approaches: 1. Collaborative filtering (implementation of word-of-mouth promotion) 2. Content-based filtering (using search keywords to determine recommendations) 3. Knowledge-based recommenders, and 4. Group recommenders (recommendations for groups). INTELLIREQ supports early requirements engineering where the major focus is to prioritize high-level requirements in software projects. INTELLIREQ automatically discerns possible dependencies among requirements. The outcome of INTELLIREQ is a consistent set of high-level requirements with corresponding effort estimations and a release plan for the implementation of the identified requirements.

Another aspect of RE is the software requirements selection challenge. It is a problem which drives the choice of the set of requirements which will be included in the next release of a software system. The optimization of this process is at the core of the Next Release Problem (NRP). NRP is an *NP hard problem*, which simultaneously manages two independent and conflicting objectives which must be simultaneously optimized: the development effort (cost), and the clients' satisfaction. This problem cannot be managed by traditional exact optimization methods. In this case, **multi-objective evolutionary algorithms** (MOEAs) are the most appropriate strategies because MOEAs tackle simultaneously several conflicting objectives without the artificial adjustments included in classical single-objective optimization methods. Further complexity is added with the Multi-Objective NRP (MONRP). When managing real instances of the NRP, the problem is explained by the requirements tackled suffering interactions. Chaves-González et al. [39] adopted a novel multi-objective Teaching Learning Based Optimization (TLBO) algorithm to solve real instances of NRP. TLBO is a **Swarm**

**Intelligence** algorithm which uses the behavior of participants in a classroom to solve the requirements selection problem. In this context, the original TLBO algorithm has been adapted to solve real instances of the problem generated from data provided by experts. The authors tested the efficacy of a TLBO scheme modified to work with the multi objective optimization problem (MOOP) addressed in the study. Various numerical and mathematical experiments were conducted and confirmed the effectiveness of the multi-objective proposal. The results of the experiments demonstrated that the multi-objective TLBO algorithm developed performs better than other researched algorithms designed to do the same task. Contrary to that, Sagrado et al. [40] proposed an application through **Ant Colony Optimization** (ACO) for NRP. The idea is to help engineers take a decision about which set of requirements has to be included in the next release. The proposed ACO system is evaluated by means of sorting through a **Genetic Algorithm** (called: NSGA-II) and a Greedy Randomized **Adaptive Search** Procedure (GRASP). Neumann [41] however, presented an enhanced technique for risk categorization of requirements, Principal Component Analysis through **Artificial Neural Networks** (ANN). This technique improved the capability to discriminate high-risk software aspects. This approach is built on the combined strengths of **Pattern Recognition** and ANN. Principal component analysis is utilized to provide means of normalizing the input data; thus, eliminating the ill effects of multi-collinearity. A neural network is used for risk determination and classification. This procedure provides the technique with capability to discriminate data-sets that include disproportionately large numbers of high-risk software modules. The combination of principal component analysis and ANNs has shown to provide significant improvements over either method by itself. Another aspect of RE, is project planning. It is a crucial activity in software projects which significantly affects the success or failure of a project. Accurate estimation of software development efforts plays a vital role in management of software projects because it can significantly affect project scheduling and planning. Underestimation of software project efforts causes delay and cost overrun, which can lead to project failure. Conversely, overestimation can also be detrimental for effective utilization of project resources. Specific past-experience of individual situations may be a good guiding factor in such situations. Vasudevan [42] presented one of the first experience-based models for software project management (in 1994). In their approach, the focus was on using concrete cases or episodes, rather than on basic principles. **Fuzzy Logic** was employed to represent case indices and fuzzy aggregation functions to evaluate cases. This provided a formal scheme to quantify the partial matches of a given problem with multiple cases in the database and to utilize these partial matches to compute an aggregated result. Cost estimation and risk assessment functions of software project management were also covered. The author discussed the basic principles of **Case-Based Reasoning** (CBR) as it provided a functional description of the proposed system with details of case representation and case evaluation strategies. CBR is an analogical reasoning method that stores and retrieves past solutions at specific episodes. A case-base is a memory bank that represents experience. In this research, the author presented a scheme for software project management that utilizes specific instances of past experience. The author employed a fuzzy representation of case indices that would lead to a partial generalization of cases to make them applicable to a class of situations. In the same year (1994), Black, Jr. [43] introduced a system for using AI in requirements management, called Systems Information Resource/Requirements Extractor (SIR/REX). The first part of SIR/REX or SIR provided support to the earliest phases of systems engineering by automatically creating linkages between certain fragments of information. SIR could function as a glossary to documents with alphabet soup in many cases, simplifying readability for readers. Source documents were thus transformed into enriched documents. The second portion of SIR/REX was a Requirements Extractor (REX), which used Al natural language analysis to extract candidate requirements. The task of obtaining the requirements expressed by a specification was usually a tedious task. However, SIR/REX initially analyzed a natural language document, and choose specific sentences with high probability of being candidate requirements. By searching for morphological root words instead of just key words, a filter could better search for terms in a document. When human behavior was compared to SIR/REX algorithms in requirement extraction efficiency, SIR/REX algorithms proved to be superior in both consistency and speed. From these practices, the quality and speed of requirements candidate selection was improved greatly.

Once all the requirements specifications are ready the further analysis has to be performed for software cost estimation. Soft computing based approach like ANN, Fuzzy Logic have been widely used. However, due to inconsistency and vagueness of software project attributes, an accurate estimation of development effort seems to be unreachable in a dataset comprised of heterogeneous projects. Bardsiri et al. [44] proposed a high-performance model, Localized Multi-Estimator (LMES), in which the process of effort estimation is localized. In LMES, software projects are labeled on underlying characteristics. Then, project clusters are investigated and the most accurate estimators are selected for each cluster. LMES is a combination of project classification and estimator selection. To find out the most effective estimators, estimator vector (EV) and solution vector (SV) are produced. Estimator performance is evaluated using many widely accepted metrics including magnitude of relative error (MRE), mean magnitude of relative error (MMRE), median magnitude of relative error (MdMRE), and percentage of the prediction (PRED). In this model, the development effort estimation is localized through assigning different estimators to different software projects. The investigation domain included five single and five hybrid effort estimation models. Three real datasets were utilized to evaluate the performance of LMES using widely accepted performance metrics. The evaluation results showed that LMES outperformed the other estimators. LMES is found to be quite flexible to handle the heterogeneous nature of software project datasets through the localization idea. Furthermore, LMES is not dependent on a particular type of estimator because it considers different types of algorithmic and non-algorithmic estimators for the estimation purpose. This makes LMES able to deal with the uncertainty that exists in the performance of estimation models.

Moosavi et al. [45] presented a new estimation model based on a combination of adaptive neuro-fuzzy inference system (ANFIS) and Satin Bower Bird Optimization Algorithm (SBO). SBO is a novel optimization algorithm proposed to adjust the components of ANFIS through applying small and reasonable changes in variables. Although ANFIS is a strong and fast method for estimation, the non-normality of software project data sets makes the estimation process challenging and complicated. To deal with this problem, Moosavi et al. suggested adjusting accurate parameters for ANFIS using meta-heuristic algorithm SBO. The proposed hybrid model is an optimized neuro-fuzzy based estimation model which is capable of producing accurate estimations in a wide range of software projects. The main role of SBO was to find the best parameters for ANFIS that reach most accurate estimates. Amongst many software effort estimation models, Estimation by Analogy (EA) is still one of the preferred techniques by software engineers because it mimics the human problem-solving approach. Accuracy of such a model depends on the characteristics of the dataset, which is subject to considerable uncertainty. To overcome this challenge, Azzeh et al. [46] proposed a new formal EA model based on the integration of Fuzzy set theory with Grey Relational Analysis (GRA). Fuzzy set theory is employed to reduce uncertainty in distance measures between two tuples at the $K^{th}$ continuous feature. GRA is a problem-solving method that is used to assess the similarity between two tuples with M features. Since some of these features are not necessarily continuous and may have nominal and ordinal scale type, aggregating different forms of similarity measures increases uncertainty in a similarity degree. Thus, the GRA is mainly used to reduce uncertainty in the distance measure between two software projects for both continuous and categorical features. Both techniques are suitable when relationship between effort and other effort drivers is complex. Figure 2 shows the software effort estimation framework of the GRA-based system.

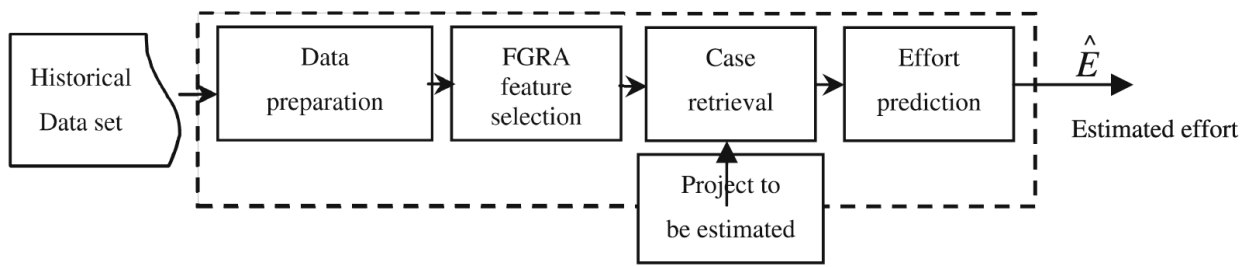

*Figure 2: A GRA Software Effort Estimation Framework [46]*

The case retrieval stage in GRA of FGRA aims to retrieve historical projects that exhibit large similarity with projects under investigation. The effort prediction stage derives final effort estimate based on the retrieved projects. In this stage, the aim is to determine the number of retrieved projects that should be involved in the effort prediction. The proposed FGRA model produced encouraging results on five publicly available datasets when compared to well-known estimation models (CBR and ANN).

Another software cost estimation product, IASCE, a prototypical expert system for estimating the cost of a proposed project in an integrated computer-aided software engineering (CASE) has been developed by Wang et al. [47]. IASCE provides support for learning, i.e. tailoring existing models, based on experience to the specific needs and characteristics of the environment. It supports multiple software cost estimation models and their corresponding metrics tractable for project management control, feedback and learning activities. In addition, IASCE provides for the establishment of project specific cost models and corporate metrics for the models, enables tracing of these models and metrics throughout the software lifecycle via feedback and post mortem evaluations, and offers a mechanism for long range improvements of software cost estimation. The basic architecture of IASCE consists of Expert System Monitor (ESM), Expert Judgment, Expert Shell1, Expert Shell2, Model Evaluation, and Model and Metrics Interpreter (MM1). Expert System Monitor (ESM) controls the execution of IASCE system modules. The Expert Judgment tool is consulting with one or more experts who use their experience and understanding of the proposed project to adjust the estimating results which were given by the expert system. Expert Shell1 contains the inference strategies and control that simulates expert model processing, it manipulates the rules in order to produce final cost estimation. Expert Shell2 plays the role of an administrator for entry of both new rules and facts. Benala et al. [48] used the combined Fuzzy C-Means (FCM) **Data Clustering** algorithm and Functional Link Artificial Neural Networks (FLANN) to achieve accurate software effort prediction. FLANN is a computationally efficient non-linear network and is capable of complex nonlinear mapping between its input and output pattern space. The non-linearity is introduced into the FLANN by passing the input pattern through a functional expansion unit. The proposed method uses three real time datasets.

ANN techniques are very popular for prediction of software development effort due to its capability to map non-linear input with output. Hota et. al, [49] explored Error Back Propagation Network (EBPN) for software development effort prediction, by tuning two algorithm specific parameters, learning rate and momentum. EBPN is a kind of NN popularly used as predictor due to its capability of mapping high dimensional data. Error back propagation algorithm is used with EBPN which is one of the most important developments in neural networks. The experimental work was carried out with WEKA open source data mining software which provides an interactive way to develop and evaluate model. EBPN was then tested with two benchmark datasets, *China* and *Maxwell*. The authors demonstrated the ability of ANN for prediction of software development effort. Further, Nassif et al. [50] carried out the research to compare four different neural network models: Multilayer Perceptron (MLP), General Regression Neural Network (GRNN), Radial Basis Function Neural Network (RBFNN), and Cascade Correlation Neural Network (CCNN) for a software development effort estimation. In their study [50] the four different neural network models: MLP, GRNN, RBFNN, and CCNN, were compared based on: (1) predictive accuracy centered on the mean absolute error criterion, (2) whether such a model tends to overestimate or underestimate, and (3) how each model classifies the importance of its inputs. Industrial datasets from the International Software Benchmarking Standards Group (ISBSG) were used to train and validate the four models. The main ISBSG dataset was filtered and then divided into five datasets based on the productivity value of each project. In this study, the performance criterion used was the mean absolute residual (MAR). Each model had four inputs: (1) software size, (2) development platform, (3) language type, and (4) resource level. The software effort was the output of the model. Results showed that the MLP and GRNN models tend to overestimate based on 80% of the datasets, followed by the RBFNN and CCNN models which tend to overestimate based on 60% of the datasets. An MLP is a feed-forward-typed artificial neural network model that has one input layer, at least one hidden layer, and one output layer. Each neuron of the input layer represents an input vector. If a network is only composed of an input layer and an output layer (no hidden layer), then the name of the network becomes perceptron. The MLP model is shown in Figure 3.

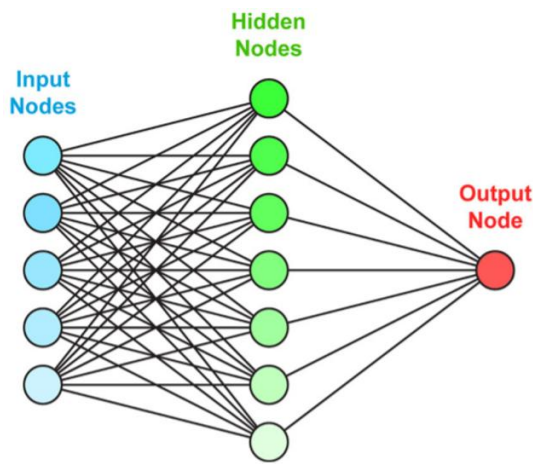

*Figure 3: Multilayer Perceptron Model (MLP) [50]*

GRNN is another type of neural network that was proposed by Specht [51]. A GRNN network applies regression on continuous output variables. A GRNN is composed of four layers as depicted in Figure 4. The first layer represents the input layer in which each predictor (aka independent variable) has a neuron. The second layer is fed from the input neurons. An RBFNN network is a feed-forward network composed of three layers: an input layer, a hidden layer with a nonlinear RBF activation function, and a linear output layer. Figure 5 shows the diagram of the RBFNN network. A CCNN network, which is also known as a self-organizing network, is composed of an input, hidden, and output layers. When the training process starts, a CCNN network is only composed of an input and output layers. Each input is connected to each output.

In the second stage, neurons are added to the hidden layer one by one. Figure 6 shows a CCNN network with one neuron.

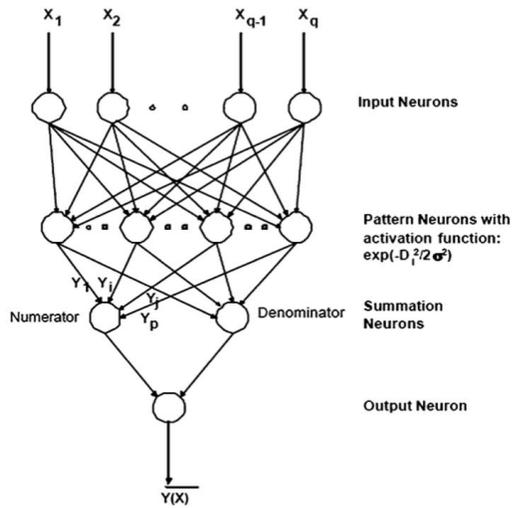

Figure 4: General Regression Neural Network (GRNN) [51]

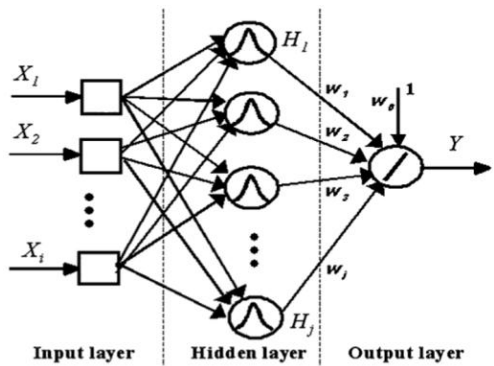

Figure 5: Radial Basis Function Neural Network (RBFNN) [50]

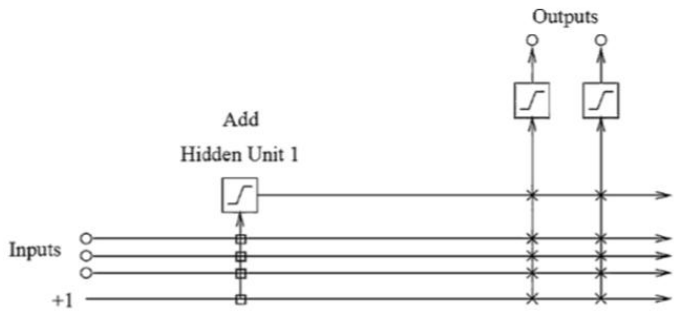

Figure 6: CCNN with One Hidden Neuron [51]

## 2.2 Software Design

Similar to Requirements, though less methods were found, AI has been applied to the Design phase as well ([52][53][54][55][56][57][58][59][60][61][62][63]). Software design is the activity of creating an engineering representation (i.e. a blueprint) of a proposed software implementation. In the design process, the software requirements document is translated into design models that define the data structures, system architecture, interfaces, and multiple components [64][65]. The software design phase usually takes only up to 6% of the software development budget, though it can drastically effect results in software development [66]. To make the design phase more efficient, AI applications have been explored. Architectural design defines the relationships amongst the major structural elements of software. Interface design describes how software elements, hardware elements, and end-users communicate with one another [65]. **Knowledge-Based Systems** (KBS) have various applications in database design. A paper from the Hungarian Academy of Science examined the function and architecture of a KBS for supporting the information system design process [67]. Multiple other AI techniques and applications are proposed in *design*. This includes a paper by Jao et al. [68], proposing a strategy based on autonomic computing technology for self-adaptations to dynamic software architectures and applying self-adaptations to autonomous agents. Figure 7 shows the architecture of autonomous agents presented in that paper.

Other papers by Taylor and Frederick [69] proposed the creation of a third-generation machine environment for computer-aided control engineering (CACE), using an **Expert System** (a synonym for KBS). The authors described the faults in the CACE software as the inspiration for applying AI to their proposed environment. That paper focused on the 'high-level requirements' for an improved CACE environment. Figure 8 shows the complete functional structure of CACE. Another paper by Dixon et al. [70] proposed an architecture focused more on evaluation and re-design, emphasizing the iterative nature of the engineering design process. The proposed architecture contains four initial functions: initial design, evaluation, acceptability, and re-design. Each of these functions are represented in the architecture by a separate knowledge source. A fifth function, control, acts as a central control module and decides which knowledge source is invoked next. A sixth function, the user interface also exists as a separate knowledge-base.

Another paper by Soria et al. [71] suggested that searching the design space can occur more efficiently through AI-based tools. The paper discuses two tools: The first tool assists exploring architectural models, while the second tool assists the refinement of design architectural models into object-oriented models (that leads to development). Additionally, a paper by Rodríguez et.al. [72] suggested solutions for a service-oriented architecture (SOA). SOA is a form of software design where services are provided to other components by application components (typically through a communication protocol). The paper offers a conceptualized analysis of AI research works that have intended to discover, compose, or develop services. The aim of the study is to classify significant works on the use of AI in determining, creating and developing Web Services.

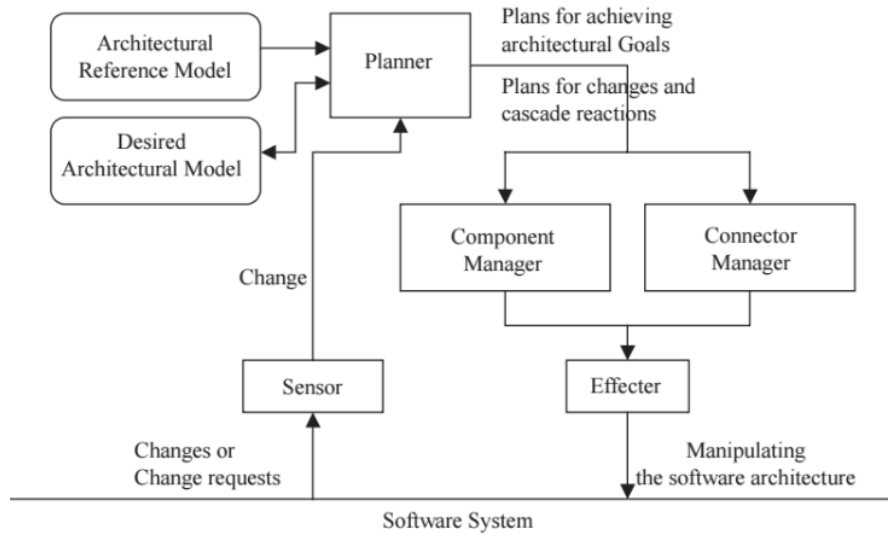

*Figure 7: The Architecture of Autonomous Agents [68]*

For AI Applications in user interfaces' (UI) design, few papers were found. A paper by William B. Rouse [73] presents a conceptual structure for the design of man-computer interfaces for online interactive systems. The design of UIs was discussed through visual information processing, and mathematical modeling of human behavior. Likely paths of study in man-computer systems were also recommended [73].

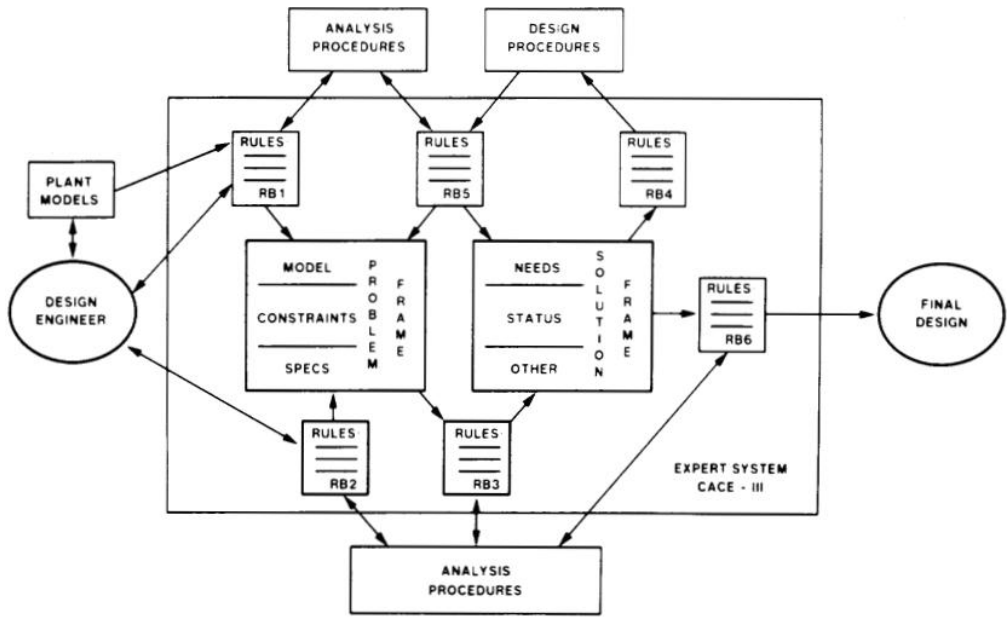

*Figure 8: Complete Functional Structure of CACE-III [69]*

Another paper was recently published by Batarseh et al. [74]. The paper presented a model to developing UIs for emergency applications through the understanding of the *context* of the user (demographics, age, nationality, and role of the user dictate how the user visualizes the UI). After design, all the decisions taken are transformed into implementation, that phase is discussed in the next section.

## 2.3 Software Development and Implementation (*Writing the Code*)

The software development phase [75], also called the implementation phase, is the phase in the lifecycle where the system is transformed from design to *production*. In this phase, the desired software components are built either from scratch or by composition. This component building process is done using the architecture or design document from the software design phase, and the requirement document from the requirements analysis phase. This phase moreover also deals with issues of quality, performance, baselines, libraries, debugging, and the end deliverable is the product itself. Different frameworks and methodologies have been followed over the years to achieve the final working product in this software development phase. The evolutionary nature of SE with long processes and stages of development required, makes the realities of requirements change by the time coding is being finished. Automated programming environments are suggested as a solution to this. This automation can be incorporated in the process of code generation, code reuse and code refactoring. AI can be applied to automate or assist developers in other programming processes such as generating functions, and data structures. The Autonomous Software Code Generation (ASCG) is an agent-oriented approach for automated code generation [76]. Insaurralde [77] also proposed Autonomous Development Process. The author proposed an approach that goes beyond the software development automation which usually involves the software synthesis from design models, and pre-defined policies and fixed rules. The author presented a self-directed development process that can make decisions to develop software. In this approach, an **Ontology-Enabled Agent** becomes the human developer by performing software development activities autonomously. Knowledge captured by the ontological database enables high level reasoning to interpret, design, and synthesize the system logic. This methodology is implemented using a graphic computer tool. The framework for Autonomous Software Code Generation is shown in Figure 9. The approach initially implements only an artificial agent -Software Developer Agent (SDA)- who starts dealing with the development of system by reading the requirements specification given as a physical configuration of the software under development. The mentioned SDA can capture this information and queries its own internal knowledge by means of a *reasoner* in order to make decisions to design the software that realizes the system logic. The system logic is built of interconnected blocks that can exchange information by receiving data from and sending data to other blocks. Following this information, the SDA can generate the software code as specified decisions to design the software that realizes the system logic. This approach moved away from conventional engineering solutions by developing software in a semi-autonomous manner, and instead deployed a purely-automated method.

Another aspect of software implementation is software reusability. Reusability is often touted as one of the most crucial processes to advancing software development productivity and quality. In reusability, developers obtain standard components whose behavior and functionality are well described and understood, and then integrate these components into a new software system. Several knowledge-based software reuse design models have been proposed in early research by Biggerstaff et.al. in [78] and [78]. The Knowledge-Based Software Reuse Environment (KBSRE) for program development supports the users to acquaint themselves with the domain application environment, to find partly matched components from the reusable component library, to comprehend the lifecycle knowledge of a component, and to decompose a component in necessary conditions. KBSRE is a system which allows the implementer to modify the knowledge base, decomposition rules, and sibling rules with minimum efforts. Wang et al. discussed three examples of such methods in their literature review [80]. The authors presented the work by Prieto et al. [81] who borrowed notions from library science and developed a multidimensional description framework of facets for classifying components. The other research works discussed were: The Programmer's Apprentice system by Waters [82], which provides the user with a knowledge-based editor (KBEmacs), and Intelligent Design Aid (IDeA), also by Biggerstaff et al. [83]. KBEmacs is based on the refinement-based software development paradigm. Other uses and techniques for using AI in coding include the Pattern Trace Identification, Detection, and Enhancement in Java (PTIDEJ). Also, **Search-Based** Software Engineering (SBSE) is used. Hewitt et al. [84] presented the idea of developing a *Programming Apprentice*, a system that is used to assist developers in writing their code as well as, establishing specifications, validating modules, answering questions about dependencies between modules, and analyzing implications of perturbations in modules and specifications. The authors proposed a method called meta-evaluation which attempts to implement the process that programmers perform to verify if their program meets specifications. Additionally, Hewitt et al. [84] described meta-evaluation as a process, which attempts to show that the contracts of an actor will always be satisfied. Here, a contract is a statement of what should happen in a program under a set of conditions.

Additionally, Software refactoring (one of the most expensive parts of software implementation) has seen many successful applications of Search Based methods (using metaheuristics [86][87][88][89][90] and others). In most of these studies, refactoring solutions were evaluated based on the use of *quality metrics*. On the other hand, Amal et al. [91] introduced the use of a neural network-based fitness functions for the problem of software refactoring. They presented a novel interactive search-based learning refactoring approach that does not require the definition of a fitness function. The software engineers could evaluate manually the suggested refactoring solutions by a Genetic Algorithm (GA) for few iterations then an Artificial Neural Network (ANN) could use these training examples to evaluate the refactoring solutions for the remaining iterations. The algorithm proposed by Amal et al. is shown in Figure 10.

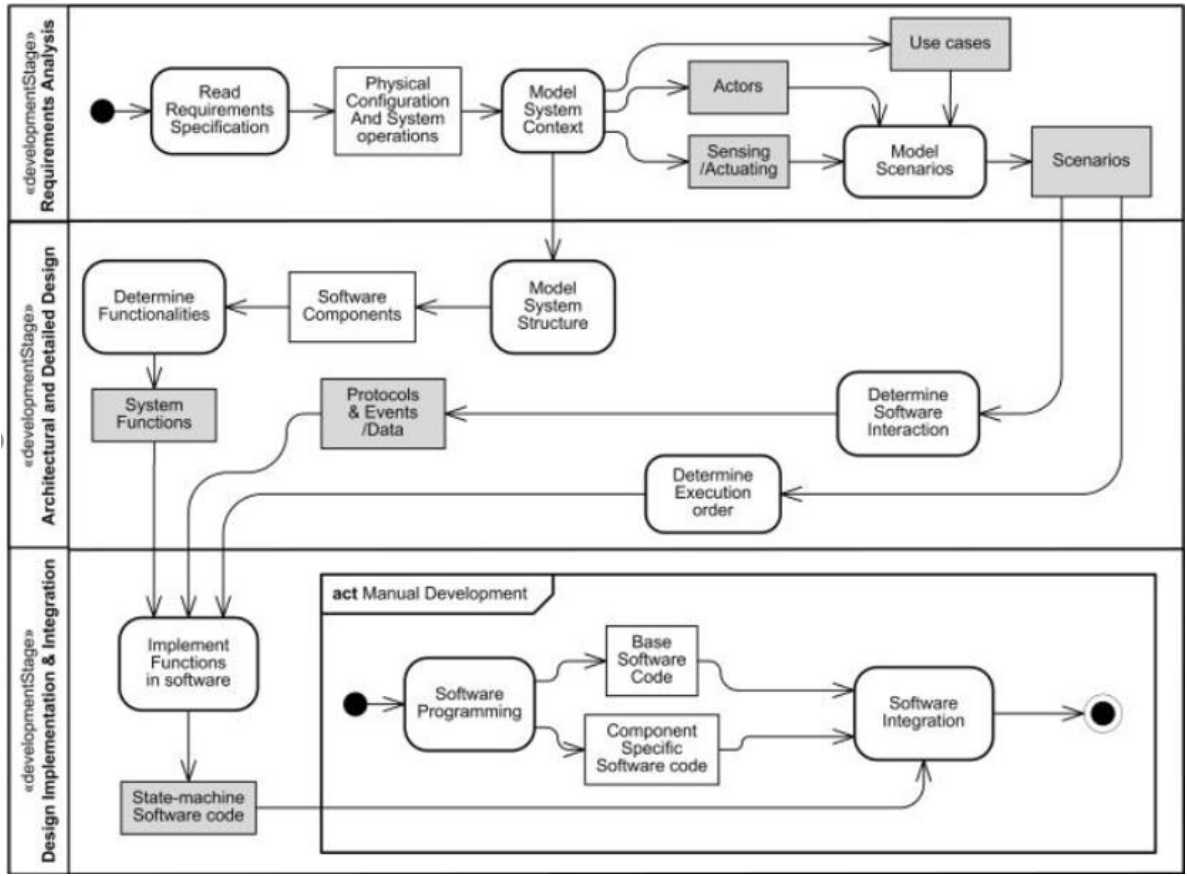

*Figure 9: Autonomous Software Code Generation framework [77]*

The method proposed in that paper the system as an input to refactor. Afterward, an exhaustive list of possible refactoring types and the number of designer's interactions during the search process are generated. The output also provides the best refactoring sequences that would improve the quality of the system. The approach is composed of two main components: the interactive component, Interactive Genetic Algorithm (IGA) and the learning module, Learning Genetic Algorithm (LGA). The algorithm starts first by executing the IGA component where the designer evaluates the refactoring solutions manually generated by GA for many iterations. The designer evaluates the feasibility, and the efficiency or quality of the suggested refactoring one by one since each refactoring solution is a sequence of refactoring operations. Thus, the designer classifies all the suggested refactoring as good or not one by one based on his preferences. After executing the IGA component for many iterations, all the evaluated solutions by the developer are considered as a training set for LGA.

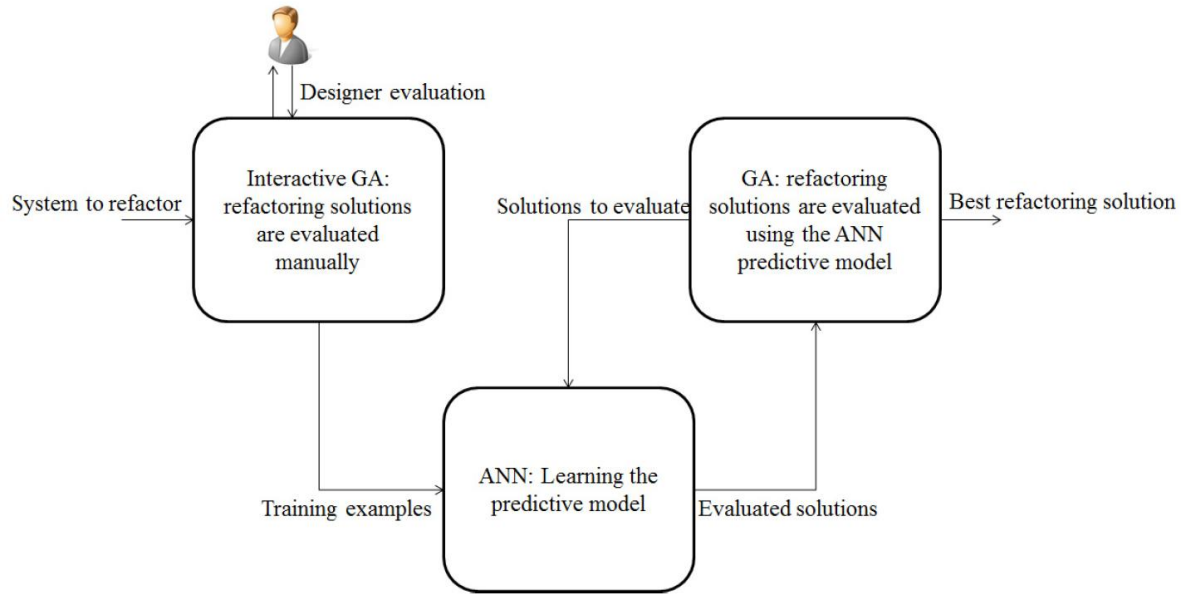

*Figure 10: IGA, LGA and ANN used for Implementation [91]*

The LGA component executes an ANN to generate a predictive model to approximate the evaluation of the refactoring solutions in the next iteration of the GA. Thus, the approach does not require the definition of a fitness function. The authors used two different validation methods: manual validation and automatic validation. These methods evaluate the efficiency of the proposed refactoring to analyzing the extent at which the proposed approach could improve the design quality and propose efficient refactoring solutions. They compared their approach to two other existing search-based refactoring approaches by Kessentini et al. [92] and Harman et al. [86]. They also assessed the performance of their proposal with the IGA technique proposed by Ghannem et al. [93], where the developer evaluates all the solutions manually. The proposed methodology in [91] required much less effort and interactions with the designer to evaluate the solutions since the ANN replace the DM after a number of iterations or interactions.

Many other automated programming environments have been proposed in the research such as Language Feature, Meta Programming, Program Browsers, and Automated Data Structuring [76]. Language Feature is a technique based on the concept of *late binding* (i.e. making data structures very flexible). In late binding, data structures are not finalized into implementation structures. Thus, quick prototypes are created, and that results in efficient codes that can be modified, and managed easily. Another important language feature is the packaging of data and procedures together in an object, thus giving rise to object-oriented programming, which is found useful in environments where codes, data structures and concepts are constantly changing [76]. LISP, one of the oldest high-level programming language is in widespread use today, provides facilities such as Meta Programming and Automated Data Structuring [76]. Meta Programming is the concept developed through NLP, which is a subfield of AI. Meta Programming is a practice in which computer programs can consider *other programs* as their data. Thus, a program can be created to read, generate, analyze or transform other programs, and even modify itself while running. Meta Programming uses automated parser generators and interpreters to generate executable LISP codes. Automated Data Structuring means going from a high-level specification of data structures to an implementation structure. When systematic changes are required to be made throughout a code, it is more 'quality-controlled' and manageable to do it through another program e.g., program update manager, than through a manual text editor for example or a traditional API.

## 2.4 Software Testing (Validation and Verification)

Software Testing is one of the major targets of AI-driven methods, many methods were found [94][95][96][97][98][99][100][101][102][103][104][105][106][107][108][108][110][111][112][113][114][115][116][117][118][119][120][121][122][123][124][125][126][127][128][129][130][131][132][133][134][135][136][137][138][139][140][141][142][143][144][145][146][147][148]. Software testing is a process of executing a program or application with the intent of finding the software defects or bugs. Software testing can also be defined as the process of Validating and Verifying (V&V) that a software program meets the business and technical requirements '*building the* **system right** (verification) *and building the* **right system** (validation)' is a popular informal definition for Testing (V&V) [4]. The testing process establishes a confidence that a product would operate as expected in a specific situation, but not ensuring that it would work well in all conditions. Testing is more than just finding bugs. The purpose of testing can be also quality assurance and reliability estimation. V&V is used to evaluate software's performance, security, usability, and robustness [149], [150]. In software testing, the relationships and interactions between software and its environment are simulated. In the next phase of selecting test scenarios, the correct test cases covering complete source code are developed and deployed, input sequences, and execution paths are also selected. This ensures that all modules of the software are adequately tested, however, this stays as one of the most interesting questions in testing – when to stop testing? After preparing and selecting test cases, they are executed and evaluated. Testers compare the outputs generated by executed test cases and the expected outputs based on defined specifications. Testers perform quantitative measurement to determine the process status by cognizing the number of faults or defects in the software [150]. Other forms of testing include field testing, graphical testing, simulation testing, and many other types.

Software testing consumes a substantial amount of total software development resources and time. It is estimated that around 40%-50% available resources and almost 50% of development time is invested in software testing [151]. AI methods which are fueled by two parts, the data and the algorithm, pose a great candidate for effective and intelligent optimization of software testing. There have been several finished works reported for applications of AI methods in software testing. One of the early proposals for using AI came through Nonnenmann and Eddy [152] in 1992. Nonnenmann et al. developed KITSS, Knowledge-Based Interactive Test Script, to ease the difficulty and cost of testing at AT&T's Bell Laboratories. KITSS works on the Private Branch Exchange (PBX) telephone switches using automated test code generation. KITSS was an automated testing system that focused on functional testing. KITSS inspected the functionality of an application without considering its internal structures. In their study, the researchers restricted KITSS to the problem of generating code tests. The project methodology for designing features included writing test cases in English, then describing the details of the external design. This was a cumbersome process, and only about 5% of the test cases were written by automation languages. Automation also presented problems, including conversion issues (test case to test script). Overall, KITSS attempted to solve such problems. Firstly, KITSS had to convert English into formal logic. Secondly, it had to extend incomplete test cases. To solve these problems, the method used a NL processor supported by a hybrid domain model and a completeness and interaction analyzer. With these formulations, the KITSS prototype system could translate test cases into automated test scripts. Eventually, KITSS turned out to be a testing process which resulted in more automation and less maintenance. The KITSS architecture is shown in Figure 11.

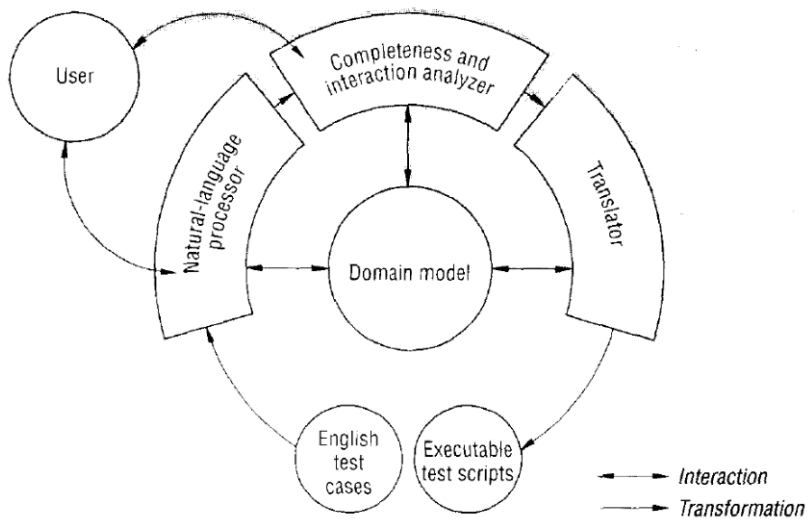

*Figure 11: The KITSS Architecture [152]*

During the mid-1990s the research on using AI techniques to improve software testing structures became more prominent. AI planners began generating test cases, providing initial states, and setting the goal as testing for correct system behavior. In [151], Annealing Genetic Algorithm (AGA) and Restricted Genetic Algorithm (RGA), two modified versions of Genetic Algorithm (GA) have been applied to testing. Sangeetha et al. [153] presented applications of GA through *white box testing*: Control Flow Testing, GA based Data Flow testing, and **Particle Swarm**

**Optimization** (PSO) based testing. Sangeetha et al. also presented GA's application in *black box testing*: GA based Functional testing. Other researchers injected AI differently, Briand et al. [154] for example, proposed a methodology to refine black-box test specifications and test suites. Suri et al. [155] presented a literature survey of all possible applications of ACO in software testing. The application of ACO at different stages of software testing process is shown in Figure 12. The highest percentage was for Test Data Generation (TDG), 57% as seen in the figure. On the other hand, there are certain limits on usage of AI in software testing which include inability to replace manual testing, being not as good at picking up contextual faults, depending on the quality test, and restricting software development during testing. Despite all the positives and negatives the research on using AI techniques in software testing has been perceived to flourish.

Regression testing is a conventional for of software testing process. Regression tests are performed to retest functionalities of software that are deployed in new versions. This much-needed process can be costly, and could be reduced using AI's ACO [156]. The ACO algorithm takes inspiration from ants finding the shortest distance from their hive to the food source. As proposed by Kire [156], this optimization technique when applied to regression test achieves the following five tasks: 1. generating path details, 2. eliminating redundant test cases, 3. generating pheromone tables, 4. selecting and prioritizing test cases, and 5. selecting the top paths with the least performance time. In this approach, selection of the test cases from the test suites reduces test efforts. Further, incorporating the test cases with prioritization using appropriate optimization algorithm leads to better and effective fault revealing. Thus, the approach results in less execution cost, less test design costs, and maximum coverage ability of codes.

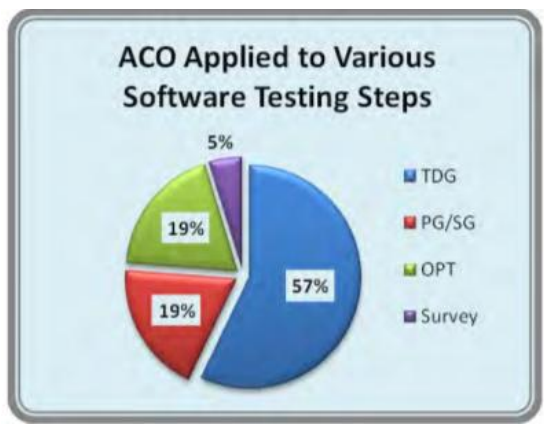

*Figure 12 Where AI Has Been Applied the Most within Testing [155]*

Moreover, GAs and fuzzy logic have been used as potential tools in selection of test cases during testing. It is very important to determine and select the right test cases. Automation of this process can significantly increases testing precision and quality [151]. In one of the review paper, Wang Jun put forward the idea of test case prioritization using genetic algorithms [157]. Ioana et al. [158] generated testing paths using test data and a GA algorithm. GA uses mechanisms inspired by biological evolution, such as reproduction, mutation, recombination, and selection. Simulated Annealing is a probabilistic technique for approximating the global optimum of a given function. Specifically, it is a *metaheuristic* to approximate global optimization in a large search space. Li et al. [159] proposed to use Unified Modeling Language (UML) Statechart diagrams and ACO to test data generation. The authors developed a tool that automatically converts a state chart diagram to a graph. The converted graph is a directed, dynamic graph in which the edges dynamically appear or disappear based on the evaluation of their guard conditions. The authors considered the problem of simultaneously dispatching a group of ants to cooperatively search a directed graph. The ants in the paradigm can sense the pheromone traces at the current vertex and its directly connected neighboring vertices, and leave pheromone traces over the vertices. The authors concluded that in this ant colony optimization algorithm, a group of ants can effectively explore the graph and generate optimal test data to satisfy test coverage requirements. Furthermore, Saraph et al. [160] discussed that analysis of inputs and outputs (I/O) to aid with testing. The authors proposed that using ANN for I/O analysis (by identifying important attributes and ranking them) can be effective. In their research [161], Xie tackled SE problems that can be solved with the synergy of AI and humans. The presented AI techniques include heuristic searching used in test generation. Traditional test generation relies on human intelligence to design test inputs with the aim of achieving high code coverage, or fault detection capability.

A Regression Tester Model, presented by Last et al. [162] and [163], introduced a full automated black box regression testing method using an Info Fuzzy Network (IFN). IFN is an approach developed for knowledge discovery and data mining. The interactions between the input and the target attributes are represented by an information theoretic connectionist network. Shahamiri et al. developed automated Oracle which can generate test cases, execute, and evaluate them based on previous version of the software under test [164]. The structure of their method is shown in Figure 13. As seen in Figure 13, Random Test Generator provides test case inputs by means of specification of system inputs. These specifications contain information about system inputs such as data type and values domain. The test bed executes these inputs on Legacy Versions (previous versions of the software under test) and receives system outputs. Next, these test cases are used to train and model IFN.

Ye et al. [165] also used ANNs to approximate the I/O behavior. Xie et al. [166] [167] proposed a cooperative testing and analysis including human-tool cooperation. That consisted of human assisted computing and human-centric computing. In human-assisted computing [166], tools have the driver seat and users provide guidance to the tools so that the tools perform the work better. In contrast, in human-centric computing [167], users have the driver seat and tools provide guidance to the users so that the users can carry out the tasks. Ye et al. [168] also proposed ANNs as an automated oracle to automatically produce outputs that compare the actual outputs with the predicted ones. Intelligent Search Agent (lSA) for optimal test sequence generation and Intelligent Test Case Optimization Agent (ITOA) for optimal test case generation have been explored [169].

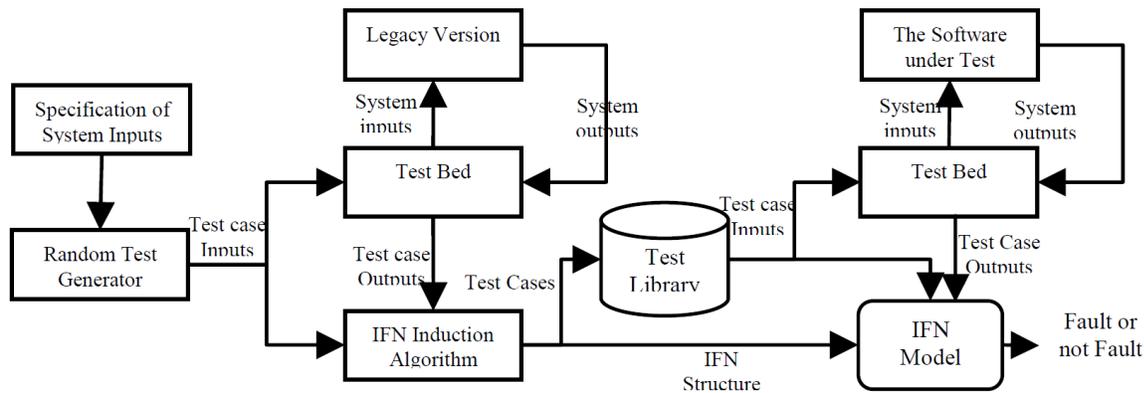

*Figure 13: IFN Based Black-Box Testing [164]*

Structural testing, is another form of tests for the internal structures or workings of a software. The fault-based testing is testing software using test data designed to demonstrate the *absence* of a set of pre-specified and frequently occurring faults. It is found that the existing testing software testing methods produce a lot of information including, input and produced output, structural coverage, mutation score, faults revealed, and many more. However, such information is not linked to functional aspects of the software. A **Machine Learning** (ML) based approach introduced by Lenz et al. [170] is used to link information derived from structural and fault-based testing to functional aspects of the program. This linked information is then used to easily test activity by connecting test results to the applications of different testing techniques. Khoshgoftaar et al. [171] suggested using Principle Component Analysis to reduce number of software metrics and extracting the most important cases. Additionally, Pendharkar [172] proposed a Software Defect Prediction Model Learning Problem (SDPMLP) where a classification model selects appropriate relevant inputs, from a set of all available inputs, and learns the classification function. The problem attempted to be solved is by finding the combination of a function f and a vector *z* such that *f (z)* has the best prediction accuracy. The solution to SDPMLP comes from identifying all values of *U*, learning *f (z)* for all values of *U* using the ML algorithm. Selecting the value(s) of $U^*$ that provide the best prediction accuracy. In this method, *z* is a vector of input attributes, *U* is a binary vector and $U^*$ provides the optimal solution. Figure 14 illustrates a general framework for solving with SDPMLP.

AI planning has also been used in testing distributed systems as well as graphical user interfaces. GUI testing is a process which is used to test an application to check if the application is working well in terms of *functional* and *non-functional* requirements. The GUI testing process involves developing a set of tasks, executing these tasks, and comparing actual results with the expected results. The technique includes detecting application's reaction to mouse events, keyboard events, and reaction of components such as buttons, dialogs, menu bars, images, toolbars and text fields towards user input. Rauf [151] proposed how Graphical User Interface (GUI) can derive benefits from the use of AI techniques. The technique includes automating test case generation so that tests are also re-generated each time GUI changes. Memon [100] proposed a method to perform GUI regression testing using **Intelligent Planning** (an AI paradigm). More generic processes can be deployed early in the process, and not through the GUI. Though software testing is adopted as a distinct phase in software development life cycle, it can be performed at *all stages* of the development process.

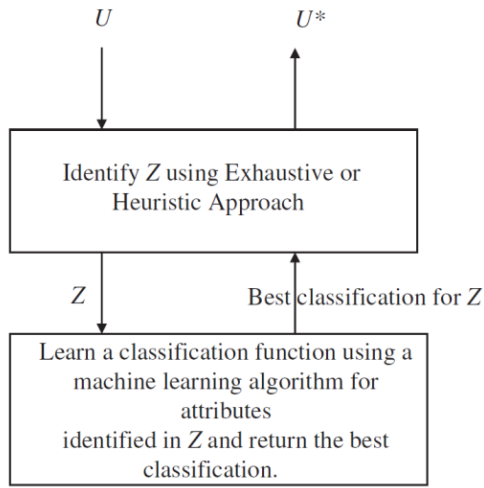

*Figure 14: A General Framework for Solving the SDPMLP [172]*

Batarseh et al. [173] introduced a novel method called Analytics-Driven Testing (ADT) to predict software failures in subsequent agile development sprints. The agile software development lifecycle is based on the concept of incremental development and iterative deliveries. *Predictive testing* is a type of testing that compares previous validation results with corresponding results of the system, iteratively. ADT predicts errors with a certain statistical confidence level by continuously measuring **Mean Time between Failures (MTBF)** for software components. ADT then uses statistical forecasting regression model for estimating where and what types of software system failures are likely to occur, this is an example of a recent method that injected **Big Data Analytics** into testing. Afzal et al. [174][175] also took inspiration from predictive testing and evaluated different techniques for predicting the number of faults such as Particle swarm optimization-based artificial neural networks (PSO-ANN), Artificial immune recognition systems (AIRS), Gene expression programming (GEP), and Multiple regressions (MR).

AI research has found its way into every stage of software testing process. Different methods and techniques are used to make the testing process intelligent and optimized. Effective software testing contributes to the delivery of reliable and quality oriented software product, and more satisfied users. Software testing thus can rely on better AI techniques. Shahamiri et al. [164] have presented a broad classification of AI and statistical methods which can be applied in different phases of automated testing. However, after evaluating all these methods, can AI answer the pressing and ongoing questions of software testing mentioned earlier? That is yet to be determined. The next section introduces AI paradigms applied to release and maintenance, the last phase in the software engineering lifecycle.

## 2.5 Software Release and Maintenance

Last, but not least, some AI paradigms were also applied to the phase of Release and Maintenance ([176][177][178][179][180][181][182][183][183][185]). Software Maintenance is defined by the Institute of Electrical and Electronics Engineers (IEEE) as the modification of a software product after delivery to correct faults, to improve performance or other attributes, or to adapt the product to a modified environment. This process generally begins with understanding what a user desires from the software product and what needs to be changed to fit what the customer wants. Feedback from users is essential for software engineers [186]. Reviews can be false positives or negatives and it is important for developers to differentiate between the true reviews and the biased ones. This issue was approached by Jindal and Liu [187] through the method of *shingling*. They grouped neighboring words and compared groups of words taken from different documents in order to search for duplicate or near duplicate reviews from different user IDs on the same product. Through their research Jindal and Liu were able to classify the types of spam review into three categories: untruthful opinions (i.e. false negative or positive reviews), brand specific reviews (reviews of a manufacturer or brand, not the product), and non-reviews (reviews that express no opinion). This could be classified as a form of **Text Mining**. Following Jindal and Liu's research, Ott et al. [188] proposed approaches for fake review detection. They found that *n-gram* based text categorization provided the best standalone detection approach. The authors of that paper created a model which used **Naïve Bayes** and **Support Vector Machines** echniques with a five-fold cross validation system in order to obtain an 86% accuracy. Following Ott et al, Sharma and Lin [189] searched for spam reviews with five criteria in mind. Firstly, they checked the product rating and customer review for inconsistency. If one was clearly positive and the other clearly negative, then the overall review is obviously untruthful. Secondly, they searched for reviews that were more about asking questions. They searched for *yes-no* question, WH-questions (why, what, where), and declarative questions. The authors used the Brill tagger to tag such sentences. Then, the model searched for reviews with all capitals. Generally, this indicates the presence of spam.

Most recently, Catal et al. [186] have proposed a model using five classifiers and a majority voting combination rule that can identify deceptive negative customer reviews. Fake review detection is done through a combination of NLP and ML in conjunction with a multiple classifiers system in order to be effective. Their research revealed that the strongest classifiers were libLinear, libSVM, SMO, RF, and J48. Catal and Guldan [186] trained their model using the dataset that Ott et al. [188] had used and had determined the accuracy rating for all of the classifiers. Catal and Guldan's model involved running a review through all five classifiers and then using a majority voting system to determine if the review was genuine or not. Their multiple classifiers model yielded an accuracy of 88.1% which is a statistically significant increase from the aforementioned 86%. However, Catal and Guldan determined that the weakness of this model was the classification cost and the complexity of the model. It is time consuming and expensive to go through the training of five different classifiers.

In the eighties, 1984 to be specific, Charles Dyer [190] researched the use of expert system in software maintainability. His paper focused on addressing the issue of planning software maintainability by applying KBS or expert systems. The reason that software maintainability can be addressed using expert systems is because maintainability can be viewed as a series of questions and depending on the answers, different support structures will be chosen. Figure 16 shows an example of a series of questions asked and then resulting conclusions provided in the bottom half of the figure. The following Figure 17 shows an example of the execution of such a system.

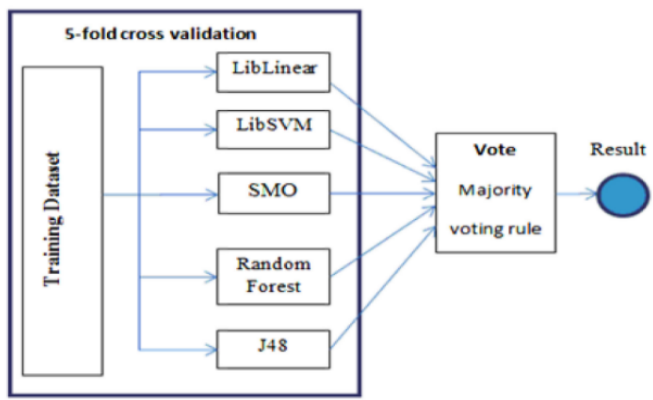

*Figure 15: Spam Detection Review Model [189]*

The advantage of using expert systems is that they offer a systematic method for capturing human judgement. In addition to this, rules can be added whenever necessary and desired by more experienced software engineers. Although it does not replace a human engineer, expert systems can most definitely serve as effective aid and support system for human judgement and decision making. However, the disadvantages of this system include that it may not be practical for sophisticated and complex systems. Such a system would require an ample amount of rules. In addition to this, there is a weakness in the fact that such an expert system exhibits no room for learning. Due to the fact that everything is very explicit, new rules will never be added automatically, and old rules will never be modified accordingly. All changes must be done manually by an engineer. That is a major drawback.

```
simple design               X X X X X X X X X X X X X X X X
complex design
high volume                    X X X X X X X X
low volume                  X                   X X X X X X X
no documentation                X X X X         X X X X
documentation available     X X         X X X         X X X
expert users                X       X X         X X           X
novice users                    X X     X             X X X X
inexpensive price           X X     X X   X     X   X X   X
expensive price                   X       X   X       X     X X
─────────────────────────────────────────────────────────────
prepare for no support
prepare for low level of support    X     X X     X X X X X       X     X
prepare for high level of support     X       X               X X     X
prepare for very high level of support    X
try for simpler design
try for lower volume
try for more documentation          X                         X
try for lower price                 X                                 X
```

*Figure 16: Maintainability Planning (as depicted from the reference) [190]*

```
-- Planning System --

PRODUCT NAME: fortplan
PRICING (EXPENSIVE OR INEXPENSIVE): inexpensive
IS VOLUME HIGH OR LOW? low
IS DESIGN SIMPLE OR COMPLICATED? complicated
IS DOCUMENTATION AVAILABLE OR IS THERE NONE? none
EXPERT OR NOVICE USERS? novice

AS PRESENTLY DESCRIBED, FORTPLAN MAY REQUIRE
A HIGH LEVEL OF SUPPORT.

IT MAY BE POSSIBLE TO SIMPLIFY THE DESIGN OF THE
PRODUCT

THE POSSIBILITY OF DISTRIBUTING A MANUAL
SHOULD BE EXPLORED.

HOW LONG WILL FIELD TEST LAST (MONTHS)? 6
IS A TECHNICAL WRITER AVAILABLE (YES OR NO)? yes
IS DEVELOPMENT ON SCHEDULE (MONTHS BEHIND)? 0
HOW MANY PAGES WOULD A MANUAL REQUIRE? 50
```

*Figure 17: Execution of Planning System (as depicted from the reference) [190]*

Following Dyer [190], Lapa et al. [191] have also investigated maintenance planning. However, their approach was to preventative maintenance planning using GA. Their research contained two goals: First to present a novel methodology for preventative maintenance policy evaluation based on a cost-reliability model. The second goal is to automatically optimize preventative maintenance policies based on the proposed methodology for system evaluation. The authors thought to use GAs because there are many parameters that need to be analyzed. GA are strong when searching for an optimum combination of options. The algorithm will analyze through the following criteria (fitness function): probability of needing a repair, cost of such a repair, typical outage times, preventative maintenance costs, impact of the maintenance reliability as a whole, and the probability of imperfect maintenance. A function that is able to evaluate the genotype, an objective function, is developed. This function takes into account the models that were built previously in order to properly evaluate the constructed genotype.

After maintenance, engineers implement additions that the user requested, or make modifications to improve the software. Lethbridge and Singer [192] have shown that searching is a major task within software maintenance. Software engineers often have to search through the source code in order to find a certain function, or perhaps to just obtain a better understanding of the software they wish to maintain. Software engineers already use a multitude of search tools like Unit 'grep' or Source Navigators. Software like these assist engineers in being able to comprehend code as well as locate parts of the software to be fixed. However, sometimes there is disconnection between what the engineer is searching for, and what is actually returned. Liu and Lethbridge [193] investigated this issue. The authors realized that the problem was that the engineer will often not query what they are looking for, which would lead to the search software not returning what the engineer desires. The authors proposed an intelligent search technique tailored to source code to assist in this endeavor. They have constructed knowledge bases to represent concepts. The knowledge base represents the hierarchies of words. During the search, the knowledge base is used to find synonyms, super-concepts, and sub-concepts of an original query. The actual process of the search starts off with many search candidates generated using several algorithms. This is done by first separating the queries into separate words and by splitting by common delimiters (e.g. spaces and underscores). Queries are then split at places where non-alphabetic characters are found prior to an upper case. A list of related terms is also stored for later evaluation. The words are then compared against a dictionary of stop words that represent all the stop words, or words to ignore. All the stop words are removed and stored in another separate list. Search queries are then performed on all of these candidates and relevant candidates remain as results. Afterwards, results must be evaluated so that an order can be produced. Functions can be created that take into account the similarity between a result and the original query string. This value is then used to weight the result and produce a list of search results. That method is shown in Figure 18.

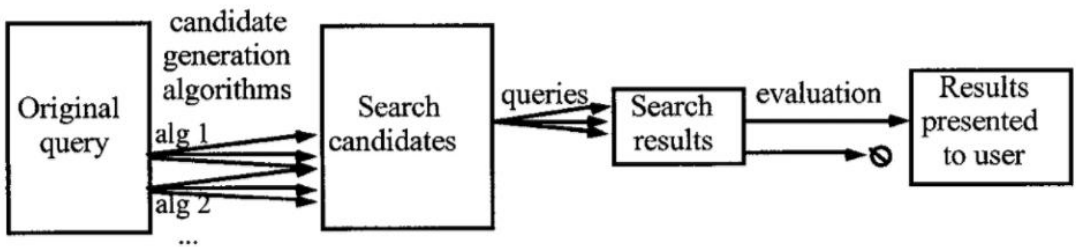

*Figure 18: Intelligent Search Procedure for Maintenance [193]*

Liu and Lethbridge have determined that their proposal for intelligent search methods has been useful to software engineers in preliminary experiments. The issue that they did find is that the response time of such a search method could be excessive at times. This is attributed to the search candidate generation. Because so many candidates are generated, it takes the system a long time to analyze and query through all of them. Mancoridis et al. [194] have developed a tool that assists in decomposing programs so that a software engineer might be able to more easily understand a particular program. This is done by *treating clustering as an optimization program*. By doing so, the authors were able to create a software that could automatically decompose a program. The tool achieves this by clustering source level modules and dependencies into sub-systems and then maps these modules and dependencies to a module dependency graph. This is done by systematically navigating through a large search space of all the possible partitions. The optimization is done by trying to maximize the value of an objective function, which when performed, represents the optimal partition of the graph. However, this is not efficient, and what BUNCH (the name of the tool presented in their paper) [194] achieved, is using more efficient search algorithms in hopes of obtaining a slightly sub-optimal partition. However, issues include not knowing where to classify a subsystem because it exists in multiple parts of the design, or event that it doesn't clearly reside in any particular part of design. Sometimes BUNCH is incorrect when compared to the intuition of an experience developer who is familiar with the code. But the overall evaluation of the tool is that it does a good job for not having any knowledge of the software design. This kind of system decomposition tool is good for software maintenance because engineers trying to alter the software now have a method in which they can take apart the software and look in particular portions of the software. They no longer have to waste time searching through parts of the code that don't effect the problem explicitly. The implementation of AI has reduced wasted time and has made the entire maintenance process more efficient. However, drawbacks that most of the presented methods share is that they might not be practical, useable or even feasible in many cases. Future work in AI-driven methods in software maintenance will have to address such issue to optimize SE overall. This and other issues are discussed in the following sections (3 and 4).

## 3. Summary of the Review

This survey chapter reviewed AI methods applied within SE between the years **1975 and 2017**. AI Methods are identified based on the definition of AI sub-fields from a book that is highly cited in the field (Artificial Intelligence: A Modern Approach) [195]. Testing and Requirements are the two phases with the highest number of contributions in the field. Scientists and researchers seem to agree that these two phases lend themselves to AI more than other. Neural Networks, Genetic Algorithms, Knowledge-Based Systems, Machine Learning and Natural Language Processing are amongst the AI paradigms applied the most. A complete list of all methods (used, reviewed or mentioned in this chapter), their associated AI paradigms, and year of publication is shown in Table 1. The method names are provided in the table to the '*best of our knowledge*'. Some papers in this list didn't have an explicit name for their method. For some others, it was a bit ambiguous to what AI paradigm was used. Please contact the authors for a comprehensive index of further references, details, and more discussions per method that were not included due to either the length of this paper, or relevance.

Table 1: A Complete List for all AI-driven Methods for SE (Ordered by Year 1975-2017)

| Name of AI-Driven Method (Paper) | AI Paradigm | SE Phase | Year |
|---|---|---|---|
| Man-computer interfaces | Bayesian Nets | Design | **1975** |
| Constructing a programming apprentice (Planner Project) | Knowledge-Based Systems | Development | **1975** |
| PSL/PSA | Automation | Release and Maintenance | **1976** |
| Informality in program specifications | Natural Language Processing | Requirements | **1977** |
| Programmer's Apprentice (PA) | Knowledge-Based Systems | Development | **1982** |
| Architecture for applying AI to design | Knowledge-Based Systems | Design | **1984** |
| Creating a third-generation machine environment for computer-aided control engineering (CACE) | Rule-Based System | Design | **1984** |
| Intelligent maintenance | Expert Systems | Release and Maintenance | **1984** |
| Classifying reusable modules | Classification | Development | **1987** |
| Code allocation for maintenance | Knowledge-Based Systems | Release and Maintenance | **1987** |
| KEE-connection | Knowledge-Based Systems | Design | **1988** |
| Project planning and control | Knowledge-Based Systems | Requirements | **1990** |
| AI applications to information systems analysis | Knowledge-Based Systems | Design | **1991** |
| LaSSIE | Knowledge-Based Systems | Development | **1991** |
| Automatic extraction of candidate requirements | Automation | Requirements | **1992** |
| IASCE: a computer-aided software design tool | Expert Systems | Requirements | **1992** |
| Predicting development faults with ANN | Neural Networks | Testing | **1992** |
| Risk assessment in software reuse | Knowledge-Based Systems | Requirements | **1993** |
| KITSS-a knowledge-based interactive test script system | Knowledge-Based Systems | Testing | **1993** |
| A software reuse environment for program development | Knowledge-Based Systems | Development | **1994** |
| Software project management | Case-Based Reasoning | Requirements | **1994** |
| SIR/REX (Systems Information Resource/Requirements Extractor) | Natural Language Processing | Requirements | **1994** |
| Early detection of program modules in the | Neural Networks | Release and | **1995** |
| maintenance phase | | Maintenance | |
| Exploring the behavior of software quality models | Neural Networks | Testing | **1995** |
| Dynamic TDG | Genetic Algorithms | Testing | **1997** |
| TCG | Planning and Heuristics | Testing | **1997** |
| A model for the design and implementation of software systems | Multi-agent AI Implementations | Design | **1998** |
| BUNCH | Clustering, Optimization Algorithms | Release and Maintenance | **1999** |
| Test-data generation | Genetic Algorithms | Testing | **1999** |
| Automatic re-engineering of software | Genetic Programming | Release and Maintenance | **2000** |
| Automatic software test data generation | Genetic Algorithms | Testing | **2001** |
| Fault allocations in the code | Search Methods | Release and Maintenance | **2002** |
| Software project effort estimation | Genetic Programming | Requirements | **2002** |
| NL-OOPS: A requirements analysis tool | Natural Language Processing | Requirements | **2002** |
| Comparison of artificial neural network and regression models | Neural Network | Requirements | **2002** |
| Software risk analysis | Neural Networks | Requirements | **2002** |
| Devising optimal integration test orders | Genetic Algorithms | Testing | **2002** |
| Using automatic test case optimization for NET components | Genetic Algorithms | Testing | **2002** |
| Predicting software development faults | Neural Networks | Testing | **2002** |
| Software cost estimation | Neural Networks | Requirements | **2003** |
| Extracting test sequences from a Markov software usage model | Ant Colony Optimization | Testing | **2003** |
| Automated test reduction | Data Analytics (Machine Learning) | Testing | **2003** |
| Test case generation and reduction | Machine Learning | Testing | **2003** |
| Intelligent software testing | Neural Networks /Data Mining | Testing | **2003** |
| Automatic software testing | Automation | Testing | **2003** |
| Automated adaptations to dynamic software architectures | Autonomous Agents | Design | **2004** |
| Software test data generation | Ant Colony Optimization | Testing | **2004** |
| Automated GUI regression testing | Planning and Heuristics | Testing | **2004** |
| Black box testing | Fuzzy Networks | Testing | **2004** |
| Automated test data generation | Data Analytics (Machine Learning) | Testing | **2004** |
| Estimating the defect content after an inspection | Machine Learning | Testing | **2004** |
| Test case generation and reduction | Neural Networks | Testing | **2004** |
| An approach to test Oracle | Neural Networks | Testing | **2004** |
| An approach for QoS-aware service composition | Genetic Algorithms | Design | **2005** |
| Adaptive fuzzy logic-based framework | Fuzzy Logic | Requirements | **2005** |
| Test sequence generation for state-based software testing | Ant Colony Optimization | Testing | **2005** |
| An approach for integrated machine fault diagnosis | Data Analytics (Machine Learning) | Testing | **2005** |
| Automatic Test Case Optimization | Genetic Algorithms | Testing | **2005** |
| A dynamic optimization strategy for evolutionary testing | Genetic Algorithms | Testing | **2005** |
| Stress testing real-time systems | Genetic Algorithms | Testing | **2005** |
| Mutation-based Testing | Genetic Algorithm (GA), Bacteriological Algorithm (BA) | Testing | **2005** |
| Software reliability estimations using dynamic weighted combinational models | Neural-networks | Design | **2006** |
| A search-based approach to improve the subsystem structure of a software system | Evolutionary Algorithm | Development | **2006** |
| Evaluating project management internal efficiencies | Fuzzy Logic | Release and Maintenance | **2006** |
| An early warning system for software quality improvements and project management | Fuzzy Logic | Release and Maintenance | **2006** |
| A model for preventive maintenance planning | Genetic Algorithms | Release and Maintenance | **2006** |
| Component selection and prioritization for the next release problem | Search Algorithms | Release and Maintenance | **2006** |
| Optimization of analogy weights | Genetic Algorithms | Requirements | **2006** |
| Capability based project scheduling | Genetic Algorithms | Requirements | **2006** |
| Software risk analysis | Neural Networks | Requirements | **2006** |
| Evolutionary unit testing of object-oriented software | Genetic Programming | Testing | **2006** |
| Automated test oracle for software testing | Neural Networks | Testing | **2006** |
| Testing XML-based multimedia software applications | Planning | Testing | **2006** |
| Pareto optimality for improving search-based refactoring | Search Algorithms | Development | **2007** |
| Bi-Objective Release Planning for Evolving Systems (BORPES) | Decision Support | Release and Maintenance | **2007** |
| Statistical Software Testing | Machine Learning | Testing | **2007** |
| Automated web service composition | Planning and Heuristics | Design | **2008** |
| Creating a Search-based tool for software maintenance | Search-Based Refactoring | Development | **2008** |
| Segmented software cost estimation models | Fuzzy Clustering | Requirements | **2008** |
| Fuzzy critical chain method in project scheduling | Genetic Algorithms | Requirements | **2008** |

| Time-cost trade off in project scheduling | Genetic Algorithms | Requirements | **2008** |
|---|---|---|---|
| Applying search based techniques requirements optimization | Search Based Techniques | Requirements | **2008** |
| Testing the effectiveness of early life cycle defect prediction | Bayesian Nets | Testing | **2008** |
| Assessment of software defect prediction techniques | Machine Learning | Testing | **2008** |
| Refining black-box test specifications and test suites | Machine Learning | Testing | **2008** |
| IntelligenTester | Multi-Agents | Testing | **2008** |
| Classifying automated testing | Neural Networks , CBR, AI Planning | Testing | **2008** |
| Recommender system for software project planning | Case-Based Reasoning | Requirements | **2009** |
| The use of a fuzzy logic-based system in cost-volume-profit analysis under uncertainty | Fuzzy Logic | Requirements | **2009** |
| Software effort estimation | Fuzzy Grey Relational Analysis | Requirements | **2009** |
| Approach for critical path definition | Fuzzy Logic | Requirements | **2009** |
| Estimate the software development effort | Fuzzy Neural Network | Requirements | **2009** |
| R-TOOL-A Natural Language based tool | Natural Language Processing | Requirements | **2009** |
| Generation of Pairwise Test Sets | Artificial Bee Colony | Testing | **2009** |
| Software Testing | Genetic Algorithms | Testing | **2009** |
| Software Testing | Genetic Algorithms | Testing | **2009** |
| Automatically finding patches | Genetic Programming | Testing | **2009** |
| Fault diagnosis and its application to rotating machinery | Neuro-Fuzzy | Testing | **2009** |
| Web service composition in cloud computing | Planning and Heuristics | Design | **2010** |
| QoS-based dynamic web service composition | Ant Colony Optimization | Design | **2010** |
| Autonomic, self-organizing service-Oriented Architecture in service ecosystem | Automation | Design | **2010** |
| Analyzing what makes software design effective | Context-Based Reasoning | Design | **2010** |
| Software reliability modeling | Genetic Programming | Design | **2010** |
| Supporting quality-driven software design | Ontologies and Planning/Automation | Design | **2010** |
| HACO algorithm for next release problem (NRP) | Hybrid Ant Colony Optimization Algorithm (HACO) | Release and Maintenance | **2010** |
| Optimized fuzzy logic based framework | Fuzzy Logic | Requirements | **2010** |
| Predicting the development effort of short scale programs | Fuzzy Logic | Requirements | **2010** |
| Search-based methods for software development effort estimation | Genetic Programming | Requirements | **2010** |
| Statistical based testing | Automated Search | Testing | **2010** |
| Presenting a regression test selection technique | Cluster Analysis | Testing | **2010** |
| Repairing GUI Test Suites | Genetic Algorithms | Testing | **2010** |
| Proposals for a Software Defect Prediction Model Learning Problem (SDPMLP) | Probabilistic Neural Network (PNN) | Testing | **2010** |
| Autonomic computing approach in service oriented architecture | Case-Based Reasoning | Design | **2011** |
| QoS-aware automatic service composition | Functional Clustering | Design | **2011** |
| An automated approach for the detection and correction of design defects | Genetic Algorithm (GA) and Genetic Programming (GP) | Development | **2011** |
| Software release planning with dependent requirements | Ant Colony Optimization | Release and Maintenance | **2011** |
| Analogy-based software effort estimation | Fuzzy Logic | Requirements | **2011** |
| Test Case Selection and Prioritization | Ant Colony Optimization | Testing | **2011** |
| GA based Test Case Prioritization Technique | Genetic Algorithms | Testing | **2011** |
| Semi-Supervised K Means (SSKM) | Machine Learning | Testing | **2011** |
| Intelligent test oracle construction for reactive systems | Machine Learning | Testing | **2011** |
| Automated choreographing of Web services | Planning and Heuristics | Design | **2012** |
| A review and classification of literature on SBSE | Search Based Optimization (SBO) | Development | **2012** |
| Software effort prediction | Fuzzy Clustering, functional link Neural Networks | Requirements | **2012** |
| Automated analysis of textual use-cases | Natural Language Processing | Requirements | **2012** |
| Generating test data for structural testing | Ant Colony Optimization | Testing | **2012** |
| Research of path-oriented test data generation | Ant Colony Optimization | Testing | **2012** |
| Automatic test data generation for software path testing | Evolutionary Algorithms | Testing | **2012** |
| Automatic generation of software test data | Hybrid Particle Swarm Optimization (PSO) | Testing | **2012** |
| Puzzle-based Automatic Testing environment (PAT) | Mutation/GE and Planning | Testing | **2012** |
| Autonomous development process for code generation | Artificial Agents | Development | **2013** |
| Using a multi-objective optimization-based approach for software refactoring | Evolutionary Algorithm | Development | **2013** |
| Model refactoring | Genetic Algorithms | Development | **2013** |
| LMES: localized multi-estimator model | Classification and Clustering | Requirements | **2013** |
| Framework for software development effort estimation | Neural Network and Fuzzy Logic | Requirements | **2013** |
| Automated software testing for application maintenance | Artificial Bee Colony | Testing | **2013** |
| Automated TDG technique for Object oriented software | Genetic Algorithms | Testing | **2013** |
| Linking software testing results | Machine Learning | Testing | **2013** |
| NSGA-III | Evolutionary Algorithm | Development | **2014** |

| Defining a fitness function for the problem of software refactoring | Genetic Algorithms | Development | **2014** |
|---|---|---|---|
| Multi-objective for requirements selection | Ant Colony Optimization | Requirements | **2014** |
| INTELLIREQ | Knowledge-Based Systems | Requirements | **2014** |
| ANFIS model | Neuro Fuzzy | Requirements | **2014** |
| An approach for the integration and test order problem | Multi-objective optimization | Testing | **2014** |
| Multi-objective teaching learning based optimization (MO-TLBO) algorithm | Evolutionary Algorithm | Requirements | **2015** |
| Predicting software development effort | Neural Networks | Requirements | **2015** |
| Performance Evaluation of software development effort estimation | Neuro-Fuzzy | Requirements | **2015** |
| Analytics-driven testing (ADT) | Data Analytics (Machine Learning) | Testing | **2015** |
| Improved estimation of software development effort | Fuzzy Analogy | Requirements | **2016** |
| Models for software development effort estimation | Neural Networks | Requirements | **2016** |
| Enhanced software effort estimation | Neural Networks | Requirements | **2016** |
| Optimized class point approach | Neuro-Fuzzy | Requirements | **2016** |
| Product review management software | Classification | Release and Maintenance | **2017** |
| Satin bowerbird optimizer: A new optimization algorithm | Neuro-Fuzzy | Requirements | **2017** |
| Automated use case diagram generation | Natural Language Processing | Requirements | **2017** |
| Context-driven UI development (CDT) | Context | Development | **2017** |

As the length of the table of Table 1 dictates, there are many AI-driven methods, however, to summarize that, the following stats are reported: Years 1975-2017, for the requirements phase, **46** major AI-driven methods were found, **19** for design, **15** for development, **68** for testing, **15** for release and maintenance. Many AI paradigms (if not most) have been applied to a certain phase of SE. Success is not limited to any of the paradigms, no correlation was found between the SE phase and the AI paradigm used. Other insights were found though, the next section provides such discussions, challenges the conventional wisdom in most of the papers in Table 1, and provides guidance to the path forward.

## 4. Insights, Dilemmas and the Path Forward

As it is established in this chapter, multiple papers have been published to make the case for the use of AI methods in SE, and to providing examples of where AI could be used, example applications included:

1. Disambiguating natural language requirements, through text mining, natural language processing and other possible 'intelligent' means.
2. Deploying 'intelligence' to the prioritization and management of requirements.
3. Using Data Analytics, ML and ANN to predict errors in software.
4. Mining through the system to eradicate any potential run-time issues.
5. Using CBR and Context to understand user requirements and ensure acceptance.
6. Using expert systems and KBS to represent the knowledge of the user, and the system logic.
7. Use Fuzzy Logic and AI Planning to predict cost and time of software.
8. Search through the code for issues using GA or ANN algorithms.

And many other ones covered in this review.

Papers used for this review have been collected following these criteria:
1- Top cited papers in the field. A search that took almost 1 week.
2- Papers by publisher, the following publishers have been tackled: IEEE, AAAI, Springer, ACM, and Elsevier.
3- Google scholar search for papers, by SE phase.
4- Our university library, all papers found there for AI methods in SE were included.

Additionally, papers that are included are ones that follow these two rules:
1- Clearly belong to one of the phases of SE (as confirmed by the authors of the paper).
2- Directly use an AI method, and is not loosely connected to AI or knowledge engineering. For example, multiple papers were found that cover knowledge engineering and management, or search-based algorithms, but those were excluded. We looked for papers that explicitly claimed a direct relationship to AI methods.

It is deemed necessary to provide an updated state of the art of AI methods applied to SE; especially with the current hype around AI. That is one of the main motivations of this chapter (although we stopped at 2017, more recent years certainly have more papers as well, we will leave that as part of ongoing work) – however, after closely evaluating the state-of-the-art, four outcomes are evident:

1- There is a general (and informal) consensus among researchers that AI methods are good candidates to be applied to SE. No paper was found that concluded that AI should be out of SE!
2- Most AI methods were applied to **one** phase of SE, thus, the structure of section 2 of this chapter. In some cases, however, AI methods were applied across multiple phases. That was loosely deployed, and most of the time it was not clear how such a method is used in the real-world. For the formal methods reviewed in the five subsections of section 2, one can notice multiple successful methods that can improve the phase.

3- The '*engineering*' process of SE lends itself to AI and to being intelligent.
4- AI has been successfully used to build software systems, and -as reviewed- across all phases. Contrary to that, when developing an intelligent agent, an AI system, or any form of system that has intelligence, very rarely are SE best practices applied. Most of the time an AI system is built, some quick prototyping/rapid development is applied. Therefore, eliminating the need for conventional SE phases. In reality, AI systems are built either incrementally, based on trial and error, based on choosing 'what works best', or other informal models. AI paradigms are used in SE, but SE paradigms are not commonly used in AI.

The four outcomes listed above lead to a number of *dilemmas*, the first one is (challenging conventional wisdom): if more AI is applied to SE, and if the process of developing software becomes completely intelligent, then wouldn't that mean that software should start building itself? And isn't that what Artificial General Intelligence (AGI) is? Most researchers and papers included in this review were chasing after reducing development cost, removing errors, and improving the process (among many other optimizations), however, none were found that aimed to inject '*general*' intelligence into the entire SE process. Therefore, there is an implicit assumption that the SE lifecycle will still be monitored by a human – which defeats the purpose of the activity (if automation aims to control the evaluation, planning, testing and other phases of the lifecycle, the human role should solely be to monitor the process, all empirical decisions should be taken by AI algorithms). This dilemma becomes part of the traditional question about the 'goodness' of AI, AI replacing human jobs, AI democratization, and other facets (scientific and philosophical) of the AI argument that are outside the context of this survey.

The second dilemma (that challenges the status quo) is: if the SE process succeeded to be completely intelligent as most of the papers 'hoped' or aimed for, *wouldn't the fields of SE and AI morph into one field*? That is a possibility because most -if not all- AI systems are depicted through software anyways. Robots without software can't do anything, intelligent agents without software can't do anything, and so on. Therefore, if the software that 'generates' the *intelligence* is designed using AI, if its developed through AI, and if it is tested with AI, then AI **becomes** the SE lifecycle itself, and the two fields merge into one (creating unified same challenges, prospects, and research communities). Does that mean the death of the conventional SE lifecycle? A dilemma worth investigating.

The third dilemma (this one is less of a dilemma and more of a quandary) is the following: most (if not all) papers found claimed that the SE process *'lends'* itself to AI. It is deemed valid that science advance when it is studied in an interdisciplinary place (the intersection of two research fields). However, that is a claim based on the notion that SE is merely a science. Most engineers, researchers and practitioners agree that SE is both a science and an **art.** That 'artistic' perspective of applying AI to SE is clearly missing from literature, and it is very vague whether that is something that AI would be able to solve or not.

Page **37** of **52**

The fourth dilemma that begs to be addressed is: as mentioned in the outcomes, no paper was found that recommends *not* using AI in SE. Most experiments presented in the papers reviewed were looking for (or comparing between) the true positives (correctly identified results) and the false positives (incorrectly identified results) of applying AI to SE. That is a major shortcoming in the overall discussion – because given that true negatives (correctly rejected) or even false negatives (incorrectly rejected) were not explicitly explored, there is still a large blind spot in this research area. Therefore, it remains unclear, whether AI should be applied further to SE. Although there are many success stories, but how many failure stories exist? That is still to be determined. However, one thing that is not ambiguous is that at this point, SE needs AI much more than AI needs SE. Errors and bad practices can hinder the progress of AI, but can't halt the process if intelligence is accomplished. Rarely has it been the case were intelligence is not accomplished due to a software error (except in some few cases; for example: driverless cars in autopilot modes could fail as such).

The goal of AI is achieving intelligence; but the goal of SE is very different, it is building a valid, verified system, on schedule, within cost and without any maintenance, or user acceptance issues. Therefore, based on the two very different premises and promises of each field (SE and AI), and all the dilemmas that this overlapping can cause, it is not a straight forward to claim the success or the need of applying AI methods onto SE. Before such a claim is made, the questions posed in this review need to be sufficiently and scientifically answered.

# References

[1] Turing, A., "Computing Machinery and Intelligence", Mind, 1950, Vol. 45, pp. 433-460
[2] Sorte, B., Joshi, P., and Jagtap, V., "Use of Artificial Intelligence in Software Development Life Cycle - A state of the art review", International Journal of Advanced Computer Engineering and Communication Technology, 2015, Vol. 4, pp. 2278-5140
[3] Rech, J., and Althoff, K., "Artificial Intelligence and Software Engineering: Status and Future Trends", KI, 2004, Vol. 18, pp. 5-11
[4] Book: Prerssman, R. and Maxim, B., "Software Engineering, A Practitioner's Approach", 8th Edition, ISBN-13: 978-0078022128.
[5] Parachuri, D., Dasa, M., Kulkarni, A., "Automated Analysis of Textual Use-Cases: Does NLP Components and Pipelines matter?", The 19th Asia-Pacific Software Engineering Conference, 2012, Vol. 1, pp. 326-329
[6] Aithal, S., and Vinay, D., "An Approach towards Automation of Requirements Analysis", Proceedings of the International Multi Conference of Engineers and Computer Scientists, 2009, Vol. I, Hong Kong
[7] Atlee, J., and Cheng, H., "Research Directions in Requirements Engineering", Proceedings of the Future of Software Engineering, 2007, pp. 285-303

[8] (unofficial report) Black, J., "A Process for Automatic Extraction of Candidate Requirements from Textual Specifications", Unpublished manuscript, 1992, to be published at GE Research 3094

[9] Agrawal, V., and Shrivastava, V., "Performance evaluation of software development effort estimation using neuro-fuzzy model", International Journal Emerging Management Technology, 2015, Vol. 4, pp. 193–199

[10] Azzeh, M., Neagu, D., Cowling, P., "Analogy-based software effort estimation using Fuzzy numbers", Journal of Systems Software, 2011, Vol. 2, pp. 270–284

[11] Cuauhtemoc, M., "A fuzzy logic model for predicting the development effort of short scale programs based upon two independent variables", Application of Software Computing, 2010, Vol. 11, pp. 724–732

[12] Idri, A., Hosni, M., and Abran, A., "Improved Estimation of Software Development Effort using Classical and Fuzzy Analogy Ensembles", Application of Software Computing, International Software Benchmark and Standard Group, 2016, pp. 990–1019

[13] Shiyna, K., and Chopra, V., "Neural Network and Fuzzy Logic Based Framework for Software Development Effort Estimation", International Journal Advanced Computer Science and Software Engineering, 2013, Vol. 3, pp. 19–24

[14] Satapathy, S. and Rath, S., "Optimized Class Point Approach for Software Effort Estimation using Adaptive Neuro-Fuzzy Inference System Model", 2016, International Journal Computer Application Technology, Vol. 54, pp. 323–333

[15] Sheenu, R., Abbas, S., and Rizwan, B., "A Hybrid Fuzzy Ann Approach for Software Effort Estimation", International Journal Foundations Computer Science (IJFCST), 2014, Vol. 4, pp. 45–56

[16] Vishal, S., and Kumar V., "Optimized Fuzzy Logic Based Framework for Effort Estimation in Software Development", International Journal Computer Science Issues (IJCSI), 2010, Vol. 7, pp. 30–38.

[17] Ahmed, M., Omolade, M., and AlGhamdi, J., "Adaptive Fuzzy Logic-Based Framework for Software Development Effort Prediction", Information and Software Technology, 2005, Vol. 47, pp. 31–48

[18] Aroba, J., Cuadrado-Gallego, J., Sicilia, N., Ramos, I., and Garcia-Barriocanal, E., "Segmented Software Cost Estimation Models Based on Fuzzy Clustering", Journal of Systems and Software, 2008, Vol. 81, pp. 1944–1950

[19] Huang, S., and Chiu, N., "Optimization of Analogy Weights by Genetic Algorithm for Software Effort Estimation", Information and Software Technology, 2006, Vol. 48, pp. 1034–1045

[20] Huang, S., Chiu, N., "Applying Fuzzy Neural Network to Estimate Software Development Effort", Journal of Applied Intelligence, 2009, Vol. 30, pp. 73–83

[21] Heiat, A., "Comparison of Artificial Neural Network and Regression Models for Estimating Software Development Effort", Information and Software Technology, 2002, Vol. 44, pp. 911–922

[22] Ferrucci, F., Gravino, C., Oliveto, R., and Sarro, F., "Genetic Programming for Effort Estimation: An Analysis of the Impact of Different Fitness Functions", Second International Symposium on Search Based Software Engineering (SSBSE '10), 2010, pp. 89–98

[23] Yang, H., and Wang, C., "Recommender System for Software Project Planning One Application of Revised CBR Algorithm", Expert Systems with Applications, 2009, Vol. 36, pp. 8938-8945

[24] Braglia, M., and Frosolini, M., "A Fuzzy Multi-Criteria Approach for Critical Path Definition", International Journal of Project Management, 2009, Vol. 27, pp. 278-91.

[25] Idri, A., Khoshgoftaar, A., and Abran, A., "Can Neural Networks Be Easily Interpreted in Software Cost Estimation?" IEEE World Congress on Computational Intelligence, 2002, Hawaii, p. 11620-1167

[26] Zhang, Y., Finkelstein, A., and Harman, M., "Search-Based Requirements Optimization: Existing Work and Challenges", International Working Conference on Requirements Engineering: Foundation for Software Quality (REFSQ), 2008, Vol. 5025, Springer LNCS, pp. 88–94

[27] Hooshyar, B., Tahmani, A., and Shenasa, M., "A Genetic Algorithm to Time-Cost Trade Off in Project Scheduling", Proceedings of the IEEE World Congress on Computational Intelligence, 2008, IEEE Computer Society, pp. 3081-3086

[28] Zhen-Yu, Z., Wei-Yang, Y., and Qian-Lei, L., "Applications of Fuzzy Critical Chain Method in Project Scheduling", Proceedings of the Fourth International Conference on Natural Computation - China, 2008, pp. 473-477

[29] Hu, Y., Chen, J., Rong, Z., Mei, L., and Xie, K., "A Neural Networks Approach for Software Risk Analysis", Proceedings of the Sixth IEEE International Conference on Data Mining Workshops, 2006, Washington DC: IEEE Computer Society, pp. 722-725

[30] Yujia, G., and Chang, C., "Capability Based Project Scheduling with Genetic Algorithms", Proceedings of the International Conference on Intelligent Agents, Web Technologies and Internet Commerce, Washington DC: IEEE Computer Society, 2006, pp. 161-175

[31] Shan, Y., McKay, R., Lokan, C., and Essam, D., "Software Project Effort Estimation using Genetic Programming", Proceedings of the IEEE International Conference on Communications, Circuits and Systems, Washington DC: IEEE Computer Society, 2002, pp. 1108-1112

[32] Boardman, J., and Marshall, G., "A Knowledge-Based Architecture for Project Planning and Control", Proceedings of the UK Conference on IT, Washington DC: IEEE Computer Society, 1990, pp. 125-132

[33]   Yuan, F. C., "The Use of a Fuzzy Logic-Based System in Cost-Volume-Profit Analysis under Uncertainty", Expert Systems with Applications, Vol. 36, 2009, pp. 1155-63

[34]   Institute of Electrical and Electronic Engineers, IEEE Standard Glossary of Software Engineering Terminology (IEEE Standard 610.12-1990). New York, NY: Institute of Electrical and Electronics Engineers, 1990.

[35]   Balzer, R., Goldman, N., and Wile, D., "Informality in Program Specifications", IEEE Transactions on Software Engineering, 1977, Vol. se-4, No. 2, pp. 94-103

[36]   Vemuri, S., Chala, S., and Fathi, M., "Automated Use Case Diagram Generation from Textual User Requirement Documents", IEEE 30th Canadian Conference on Electrical and Computer Engineering (CCECE), 2017, Vol. 95, pp. 1182-1190

[37]   Garigliano, R., and Mich, L., "NL-OOPS: A Requirements Analysis Tool based on Natural Language Processing", Conference on Data Mining, 2002, Vol. 3, pp. 1182-1190

[38]   Ninaus, G., Felfernig, A., Stettinger, M., Reiterer, S., Leitner, G., Weninger, L., and Schanil, W., "INTELLIREQ: Intelligent Techniques for Software Requirements Engineering", Frontiers in Artificial Intelligence and Applications, 2014, pp. 1-6

[39]   Chaves-González, J., Pérez-Toledano, M., and Navasa, A., "Teaching Learning based Optimization with Pareto Tournament for the Multi-Objective Software Requirements Selection", Engineering Applications of Artificial Intelligence, 2015, Vol. 43, pp. 89-101

[40]   Sagrado, J., Del Aguila, I., and Orellana, F., "Multi-Objective Ant Colony Optimization for Requirements Selection", Journal Empirical Software Engineering, 2014, Vol .1, pp. 1–34

[41]   Neumann, D., "An Enhanced Neural Network Technique for Software Risk Analysis", IEEE Transactions on Software Engineering, 2002, Vol. 28, pp. 904–912

[42]   Vasudevan, C., "An Experience-Based Approach to Software Project Management", Proceedings of the Sixth International Conference on Tools with Artificial Intelligence, 1994, pp. 624-630

[43]   Emmett Black, J., "AI Assistance for Requirements Management", Journal of Concurrent Engineering: Research and Applications, 1994, pp. 190-191

[44]   Bardsiri, V., Jawawi, D., Bardsiri, A., and Khatibi, E., "LMES: A Localized Multiestimator Model to Estimate Software Development Effort", Engineering Applications of Artificial Intelligence, 2013, Vol. 26, pp. 2624-2640

[45]   Seyyed, H., Moosavi, S., and Bardsiri, V., "Satin Bowerbird Optimizer: A New Optimization Algorithm to Optimize ANFIS for Software Development Effort Estimation", Engineering Applications of Artificial Intelligence, 2017, pp. 1-15

[46]   Azzeh, M., Neagu, and D., Cowling, P., "Fuzzy Grey Relational Analysis for Software Effort Estimation", 2009, Journal of Empirical Software Engineering, Vol. 15, pp. 60–90

[47]   Wang, S., and Kountanis, D., "An Intelligent Assistant to Software Cost Estimation", Proceedings of the IEEE International Conference on Tools with AI, 1992, pp. 1114-1176

[48] Benala, T., Mall, R., Dehuri, S., and Prasanthi, V., "Software Effort Prediction using Fuzzy Clustering and Functional Link Artificial Neural Networks", Panigrahi, B., Das, S., Suganthan, P., Nanda, P., Swarm, Evolutionary, and Memetic Computing, Springer, Berlin Heidelberg, 2012, pp. 124–132

[49] Hota, H., Shukla, R., and Singhai, S., "Predicting Software Development Effort Using Tuned Artificial Neural Network", Jain, L., Computational Intelligence in Data Mining – Proceedings of the International Conference on CIDM, Springer India, 2015, pp. 195–203

[50] Nassif, A., Azzeh, M., Capretz, L., and Ho, D., "Neural Network Models for Software Development Effort Estimation: A Comparative Study", 2016, Neural Computing Applications, Vol. 27, pp. 2369–2381

[51] Specht, D., "A General Regression Neural Network", IEEE Transactions on Neural Networks. Vol, 2, pp. 568-576

[52] Bhakti, M., and Abdullah, A., "Autonomic Computing Approach in Service Oriented Architecture", Proceedings of the IEEE Symposium on Computers & Informatics, 2011, pp. 231–236

[53] Bhakti, M., Abdullah, A., and Jung, L., "Autonomic, Self-Organizing Service Oriented Architecture in Service Ecosystem", Proceedings of the 4th IEEE International Conference on Digital Ecosystems and Technologies, 2010, pp. 153–158

[54] Canfora, G., Di Penta, M., Esposito, R., and Villani, M., "An Approach for QoS-Aware Service Composition-based on Genetic Algorithms", Proceedings of the ACM Conference on Genetic and Evolutionary Computation, 2005, pp. 1069–1075

[55] El-Falou, M., Bouzid, M., Mouaddib A., and Vidal, T., "Automated Web Service Composition using Extended Representation of Planning Domain", Proceedings of the IEEE International Conference on Web services, 2008, pp. 762–763

[56] Zou, G., Chen, Y., Xu, Y., Huang, R., and Xiang, Y., "Towards Automated Choreographing of Web Services using Planning", Proceedings of the AAAI Conference on Artificial Intelligence, 2012, pp. 383-404

[57] Huang, R. and Xu, Y., "AI Planning and Combinatorial Optimization for Web Service Composition in Cloud Computing", Proceedings of the International Conference on Cloud Computing and Virtualization, 2010, pp. 1–8

[58] Wagner, F., Ishikawa, F., and Honiden, S., "QoS-Aware Automatic Service Composition by Applying Functional Clustering", Proceedings of the IEEE International Conference on Web services (ICWS), 2011, pp. 89–96

[59] Zhang, W., Chang, C., Feng, T., and Jiang, H., "QoS-Based Dynamic Web Service Composition with Ant Colony Optimization", Proceedings of the 34th Annual Computer Software and Applications Conference, 2010, pp. 493–502

[60] Abarnabel R., Tou F., and Gilbert V., "KEE-Connection: A Bridge Between Databases and Knowledge Bases", AI Tools and Techniques, 1988, pp. 289-322

[61]    O. Raih, "A Survey on Ssearch–based Software Design", Computer Science Review, 2010, Vol. 4, No. 4, pp. 203–249

[62]    Costa, E., Pozo, A., and Vergilio, S., "A Genetic Programming Approach for Software Reliability Modeling", IEEE Transactions on Reliability, 2010, Vol. 59, No. 1, pp. 222–230

[63]    Su, Y., and Huang, C., "Neural Network-based Approaches for Software Reliability Estimation using Dynamic Weighted Combinational Models", Journal of Systems and Software, 2007, Vol. 80, pp. 606-615

[64]    Oliver, D., "Automated Optimization of Systems Architectures for Performance", Systems Engineering in the Workplace: Proceedings of the Third Annual International Symposium of the National Council on Systems Engineering, 1993, pp. 259-266

[65]    Pressman, R., "Software Engineering: A Practitioner's Approach", Pressman and Associates, 2014

[66]    Schach, S., "Object-Oriented and Classical Software Engineering", McGraw-Hill, 2002.

[67]    Molnar, B., and Frigo, J., "Application of AI in Software and Information Engineering", Engineering Application Artificial Intelligence, 1991, Vol. 4, No. 6, pp. 439-443

[68]    Jiao, W., and Mei, H., "Automated Adaptations to Dynamic Software Architectures by using Autonomous Agents", Engineering Applications of Artificial Intelligence, 2004, Vol. 17, pp. 749-770

[69]    Taylor, J., and Fredrick, D., "An Expert System Architecture for Computer-Aided Control Engineering", IEEE, 1984, pp.1795-1805

[70]    Cohen, P., Dixon, J., and Simmons, K., "An Architecture for Application of Artificial Intelligence to Design", 21st Design Automation Conference IEEE, 1984, pp. 388-391

[71]    Soria, A., Andres Diaz-Pace, J., Bass, L., Bachmann, F., and Campo, M., "Supporting Quality-Driven Software Design through Intelligent Assistants", DOI: 10.4018/978-1-60566-758-4, Chapter 10, pp. 182

[72]    Rodríguez, G., Soria, A., and Campo, M., "Artificial intelligence in Service-Oriented Software Design", Engineering Applications of Artificial Intelligence, 2016, pp. 86-104

[73]    William B. Rouse, "Design of Man-Computer interfaces for On-line Interactive Systems", Proceedings of The IEEE, 1975, Vol. 63, pp. 847-857

[74]    Batarseh, F., and Pithidia, J., "Context-Aware User Interfaces for Intelligent Emergency Applications", Proceedings of the International and Interdisciplinary Conference on Modeling and Using Context, 2017, pp 359-369

[75]    Devanbu, P., Brachman, R., Selfridge, P., and Ballard, B., "LaSSIE: a Knowledge-Based Software Information System", IEEE proceedings 12[th] international conference on Software Engineering, 1991, pp. 249-261

[76]    K. Hema Shankari and Dr. R. Thirumalaiselvi, "A Survey on Using Artificial Intelligence Techniques in the Software Development Process", International Journal of Engineering Research and Applications, 2014, Vol. 4, pp.24-33

[77]   Insaurralde Carlos C., "Software Programmed by Artificial Agents: Toward an Autonomous Development Process for Code Generation", IEEE International Conference on Systems, Man, and Cybernetics, Pg. No. 3294-3299, 2013.

[78]   Biggerstaff, T., and Perlis, A., "Software Reusability", Volume I, Reading, Mass., Addison Wesley, 1989

[79]   Biggerstaff, T., and Perlis, A., "Software Reusability", Volume II, Reading, Mass., Addison Wesley, 1989

[80]   Wang, P., and Shiva, S., "A Knowledge-Based Software Reuse Environment for Program Development", IEEE, 1994

[81]   Prieto-Diaz, R., "Classification of Reusable Modules", IEEE Transactions on Software Engineering, 1987, Vol. 4, No. 1, pp 6-16

[82]   Waters, R., "The Programmer's Apprentice: Knowledge-Based Program Editing", IEEE Transactions of Software Engineering, 1982, Vol. 8, No. 1, pp 1-12

[83]   Biggerstaff, T., Richter. C., "Reusability Framework, Assessment, and Directions", IEEE Software, 1987, Vol. 4, No. 1, pp 41-49

[84]   Hewitt, C., and Smith, B., "Towards a Programming Apprentice", IEEE Transactions on Software Engineering, 1975, Vol. SE-l, No.1, pp. 26-45

[85]   Harman, M., Mansouri, S., and Zhang, Y., "Search-based Software Engineering: Trends, Techniques and Applications", ACM Computing Surveys, 2012, Vol. 45, pp.11

[86]   Harman, M., and Tratt, L., "Pareto Optimal Search Based Refactoring at the Design Level", Proceedings of the Genetic and Evolutionary Computation Conference – GECCO, 2007, pp. 1106–1113

[87]   O'Keeffe, M., and Cinnéide, M., "Search-based Refactoring for Software Maintenance", Journal of Systems and Software, 2008, Vol. 81, pp. 502–516

[88]   Mkaouer, W., Kessentini, M., Bechikh, S., and Deb, K., "High dimensional Search-based Software Engineering: Finding Tradeoffs Among 15 Objectives for Automating Software Refactoring using NSGA-III", Proceedings of the Genetic and Evolutionary Computation Conference, 2014

[89]   Seng, O., Stammel, J., and Burkhart, D., "Search-based Determination of Refactoring for Improving the Class Structure of Object-oriented Systems", Proceedings of the Genetic and Evolutionary Computation Conference, 2006, pp. 1909–1916

[90]   Ouni, A., Kessentini, M., Sahraoui, H., and Hamdi, M. S., "The use of development history in software refactoring using a multi-objective", Proceedings of the Genetic and Evolutionary Computation Conference, 2003, pp. 1461–1468

[91]   Amal, B., Kessentini, M., Bechikh, S., Dea, J., and Said, L.B., "On the use of machine learning and search-based software engineering for ill-defined fitness function: a case study on software refactoring", Proceedings of the 6th International Symposium on Search-Based Software Engineering (SSBSE '14), 2014

[92]  Kessentini, M., Kessentini, W., Sahraoui, H., Boukadoum, and M., Ouni, A., "Design defects detection and correction by example", Proceedings of the IEEE International Conference on Program Comprehension, 2011, pp. 81–90

[93]  Ghannem, A., El Boussaidi, G., and Kessentini, M., "Model refactoring using interactive genetic algorithm", Ruhe, G., Zhang, Y., SSBSE 2013. LNCS, 2013, Vol. 8084, pp. 96–110

[94]  Last M., Kandel A., and Bunke H., "Artificial Intelligence Methods in Software Testing", 2003, World Scientific Publishing Company, Incorporated, ISBN-13: 9789812388544

[95]  Zhang, S., Mathew, J., Ma, L. and Sun, Y., "Best Basis-based Intelligent Machine Fault Diagnosis", Mechanical Systems and Signal Processing, Vol. 19 No. 2, 2005, pp. 357-370

[96]  Challagulla, V., Bastani, F., Yen, I., and Paul, R., "Empirical Assessment of Machine Learning based Software Defect Prediction Techniques", International Journal on Artificial Intelligence Tools, 2008, Vol. 17, No. 2, pp. 389–400

[97]  Poulding, S., and Clark, J., "Efficient Software Verification: Statistical Testing using Automated Search", IEEE Transactions on Software Engineering, 2010, Vol. 36, No. 6, pp. 763–777

[98]  Weimer, W., Nguyen, T., Goues, C., and Forrest, S., "Automatically Finding Patches using Genetic Programming", International Conference on Software Engineering (ICSE 2009), Vancouver, Canada, 2009, pp. 364–374

[99]  Chan, W., Cheng, M., Cheung, S., Tse, T., "Automatic Goal-oriented Classification of Failure Behaviors for Testing XML-based Multimedia Software Applications: An Experimental Case Study", 2006, The Journal of Systems and Software, Vol. 79, pp. 602–612

[100] Memon, A., "Automated GUI Regression Testing using AI Planning", World Scientific, Artificial Intelligence Methods in Software Testing, 2004, Vol. 56, pp. 51-100

[101] Last, M. and Freidman, M., "Black-Box Testing with InfoFuzzy Networks", World Scientific, Artificial Intelligence Methods in Software Testing, 2004, pp. 21-50

[102] Saraph, P., Last, M. and Kandell, A. "Test Case Generation and Reduction by Automated Input-output Analysis", Institute of Electrical and Electronics Engineers Conference, 2003

[103] Saraph, P., Kandel, A. and Last, M., "Test Case Generation and Reduction with Artificial Neural Networks", World Scientific, 2004, pp. 101-132

[104] Khoshgoftaar, T. M., Pandya, A. S. and More, H., "Neural Network Approach for Predicting Software Development Faults", Proceedings Third International Symposium on Software Reliability Engineering, 1992

[105] Khoshgoftaar, T. M., Szabo, R. M. and Guasti, P. J., "Exploring the Behavior of Neural Network Software Quality Models", Software Engineering Journal, 1995, Vol. 10, pp. 89-96

[106] Baudry, B., "From genetic to bacteriological algorithms for mutation-based testing", Software Testing, Verification & Reliability, 2005, pp 73-96

[107]   Baudry, B., "Automatic Test Case Optimization: A Bacteriologic Algorithm", IEEE Software, Published by the IEEE Computer Society, 2005, pp. 76-82

[108]   V. Mohan, D. Jeya Mala "IntelligenTester -Test Sequence Optimization framework using Multi-Agents", Journal of Computers, 2008

[109]   Last, M., Kandel, A, and Bunke, H., "Artificial Intelligence Methods in Software Testing", Series in Machine Perception and Artificial Intelligence, 2004 Vol. 56

[110]   Huang, S., Cohen, M., Memon, A., "Repairing GUI Test Suites Using a Genetic Algorithm", International Conference on Software Testing, Verification and Validation, 2010.

[111]   Michael, C., McGraw, G., Schatz, M., "Generating Software Test Data by Evolution", IEEE Transactions on Software Engineering, 2001, Vol. 27, No.12, pp.1085-1110

[112]   Xie, X., Xu, B., Nie, C., Shi, L., and Xu, L., "Configuration Strategies for Evolutionary Testing", Proceedings of the 29th Annual international Computer Software and Applications Conference COMPSAC, IEEE Computer Society, 2005, Vol. 2.

[113]   Chen, N., and Kim, S., "Puzzle-based Automatic Testing: Bringing Humans into the Loop by Solving Puzzles, Proceedings of ASE, 2012, pp. 140–149

[114]   Doerner, K., Gutjahr, W., "Extracting Test Sequences from a Markov Software Usage Model by ACO", LNCS, Vol. 2724, 2003, pp. 2465-2476

[115]   Li, H., and Peng Lam, C., "Software Test Data Generation Using Ant Colony Optimization", Transactions on Engineering, Computing and Technology, 2005, Vol. 1, No. 1, 137-141

[116]   Li, H., and Peng Lam, C., "An Ant Colony Optimization Approach to Test Sequence Generation for State-Based Software Testing", Proceedings of the Fifth International Conference on Quality Software (QSIC'05), 2005, Vol. 5, pp. 255-264

[117]   Srivastava, P., Kim, T., "Application of Genetic Algorithm in Software Testing", International Journal of Software Engineering and Its Applications, 2009, Vol. 3, No.4, pp. 87-97

[118]   Michael, C., McGraw, G., Schatz, M., and Walton, C., "Genetic Algorithm for Dynamic Test Data Generations", Proceedings of 12th IEEE International Conference Automated Software Engineering, 1997, pp. 307-308

[119]   Diaz, E., Tuya, J., and Blanco, R., "Automatic Software Testing Using a Metaheuristic Technique Based on Tabu Search", Proceedings 18th IEEE International Conference on Automated Software Engineering, 2003, pp. 301-313

[120]   McCaffrey, J., "Generation of Pairwise Test Sets using a Simulated Bee Colony Algorithm", IEEE International Conference on Information Reuse and Integration, 2009, pp. 115-119

[121]   Mao, C., YuXinxin, C., Jifu, C., "Generating Test Data for Structural Testing Based on Ant Colony Optimization", 12th International Conference on Quality Software, 2012, pp. 98 – 101

[122]   Rui, D., Xianbin, F., Shuping, L., Hongbin, D., "Automatic Generation of Software Test Data Based on Hybrid Particle Swarm Genetic Algorithm", IEEE Symposium on Electrical and Electronics Engineering (EEESYM), 2012, pp. 670 – 673

[123]   Kumar, G., and Kumar, R., "Improving GA based Automated Test Data Generation Technique for Object Oriented Software", IEEE International Advance Computing Conference (IACC), 2013, pp. 249-253

[124]   Alzubaidy, L., and Alhafid, B., "Proposed Software Testing Using Intelligent Techniques (Intelligent Water Drop (IWD) and Ant Colony Optimization Algorithm (ACO))", Semantic Scholar, 2013

[125]   Wang, J., Zhuang, Y., Jianyun, C., "Test Case Prioritization Technique based on Genetic Algorithm", International Conference on Internet Computing and Information Services, 2011, pp. 173 – 175

[126]   Karnaveland, K. and Santhoshkumar, J., "Automated Software Testing for Application Maintenance by using Bee Colony Optimization algorithms (BCO)", International Conference on Information Communication and Embedded Systems (ICICES), 2013, pp. 327 – 330

[127]   Ioana, L., Augustin, C., Lucia, V., "Automatic Test Data Generation for Software Path Testing using Evolutionary Algorithms", International Conference on Emerging Intelligent Data and Web Technologies, 2012, pp. 1-8

[128]   Bharti, S., Shweta, S., "Implementing Ant Colony Optimization for Test Case Selection and Prioritization", International Journal on Computer Science and Engineering (IJCSE), 2011, pp. 1924-1932

[129]   Minjie, Y., "The Research of Path-oriented Test Data Generation Based on a Mixed Ant Colony System Algorithm and Genetic Algorithm", International Conference on Wireless Communications, Networking and Mobile Computing, 2012, pp. 1-4

[130]   Baskiotis, N., Sebag, M., Gaudel, M.-C., Gouraud, S., "A Machine Learning Approach for Statistical Software Testing", International Joint Conference on Artificial Intelligence, 2011, pp. 2274–2279

[131]   Briand, L., "Novel Applications of Machine Learning in Software Testing", International Conference on Software Quality, 2008, pp. 1–8

[132]   Chen, S., Chen, Z., Zhao, Z., Xu, B., Feng, Y., "Using Semi-supervised Clustering to Improve Regression Test Selection Techniques, IEEE International Conference on Software Testing, Verification and Validation, 2008, pp. 1–10

[133]   Last, M., and Kandel, A., "Automated test reduction using an Info-Fuzzy network", Software Engineering with Computational Intelligence, 2008, Kluwer Academic Publishers, pp. 235–258

[134]   Noorian, M., Bagheri, E., Du, W., "Machine learning-based software testing: towards a classification framework", Software Engineering and Knowledge Engineering (SEKE), 2011, pp. 225–229

[135] Zhang, C., Chen, Z., Zhao, Z., Yan, S., Zhang, J., Xu, B., "An Improved Regression Test Selection Technique by Clustering Execution Profiles, International Conference on Software Quality, 2008, pp. 171–179

[136] Wang, F., Yao, L., Wu, J., "Intelligent Test Oracle Construction for Reactive Systems without Explicit Specifications", Ninth IEEE International Conference on Dependable, Autonomic and Secure Computing (DASC), 2011, pp. 89–96

[137] Rauf A., and Alanazi N., "Using Artificial Intelligence to Automatically Test GUI", IEEE, 2014, pp. 3-5

[138] Padberg F., Ragg T., and Schoknecht R., "Using Machine Learning for Estimating the Defect Content After an Inspection", IEEE Transactions on Software Engineering, Vol. 30, No. 1, 2004, pp. 17-28

[139] Wappler, S., and Wegener, J., "Evolutionary Unit Testing of Object-oriented Software using Strongly-typed Genetic Programming", In Proceedings of the Eighth Annual Conference on Genetic and Evolutionary Computation ACM, 2006, pp. 1925-1932

[140] Briand, L., Labiche, Y., and Shousha, M., "Stress Testing Real-time Systems with Genetic Algorithms, Proceedings of the Conference on Genetic and Evolutionary Computation, 2005, pp. 1021- 1028

[141] Gupta, M., Bastani, F., Khan, L., and Yen, I., "Automated test data generation using MEA-graph planning", In Proceedings of the Sixteenth IEEE Conference on Tools with Artificial Intelligence, 2004, pp. 174-182

[142] Baudry, B., Fleurey, F., Jezequel, J. M., and Traon, L., "Automatic Test Case Optimization using a Bacteriological Adaptation Model: Application to. NET components", Proceedings of the Seventeenth IEEE International Conference on Automated Software Engineering, 2003, pp.253-256

[143] Baudry, B., Fleurey, F., Jezequel, J. M., & Le Traon, Y., "Genes and Bacteria for Automatic Test Cases Optimization in the .Net Environment", In Proceedings of the Thirteenth International Symposium on Software Reliability Engineering (ISSRE"02), 2002, pp. 195- 206

[144] Briand, L. C., Feng, J., & Labiche, Y., "Using Genetic Algorithms and Coupling Measures to Devise Optimal Integration Test Orders", Proceedings of the Fourteenth International Conference on Software engineering and knowledge engineering, 2002, pp. 43-50

[145] Thwin, T., and Quah, T. S., "Application of Neural Network for Predicting Software Development Faults using Object Oriented Design Metrics" In Proceedings of the Ninth International Conference on Neural Information Processing, 2002, pp. 2312-2316

[146] Assunção, W., Colanzi, T., Vergilio, S.R., Pozo, A., "A Multi-Objective Optimization Approach for the Integration and Test Order Problem", Information Scientific, 2004, pp. 119–139

[147]   Zio, E. and Gola, G. "A Neuro-fuzzy Technique for Fault Diagnosis and its Application to Rotating Machinery", Reliability Engineering and System Safety, 2004, Vol. 94 No. 1, pp. 78-88

[148]   Fenton, N., Neil, M., Marsh, W., Hearty, P., Radlinski, L., and Krause, P., "On the Effectiveness of Early Lifecycle Defect Prediction with Bayesian Nets", 2008, Empirical Software Engineering, Vol. 13, No. 5, pp. 499–537

[149]   Bhateja, N., "Various Artificial Intelligence Approaches in Field of Software Testing", International Journal of Computer Science and Mobile Computing, 2016, Vol. 5, Issue 5, pg. 278-280

[150]   Whittaker, J. A., "What is Software Testing? And Why is it so Hard? Software", IEEE, 17, 2000, pp. 70-79

[151]   Abdul Rauf and Mohammad N. Alanazi, "Using Artificial Intelligence to Automatically Test GUI", The 9th International Conference on Computer Science & Education (ICCSE 2014), 2014

[152]   Nonnenmann, U., and Eddy, J., "Software Testing with KITSS", Conference on Artificial Intelligence for Applications, 1993, pp. 25-30

[153]   Sangeetha, V., and Ramasundaram, T., "Application of Genetic Algorithms in Software Testing Techniques", International Journal of Advanced Research in Computer and Communication Engineering, 2009, Vol. 5, Issue. 10, pp. 87-97

[154]   Briand, L.C., Labiche, Y., Bawar, Z., "Using Machine Learning to Refine Black-box Test Specifications and Test Suites", International Conference on Software Quality, 2008, pp. 135–144.

[155]   Bharti, S., Shweta, S., "Literature Survey of Ant Colony Optimization in Software Testing", CSI Sixth International Conference on Software Engineering (CONSEG), 2012, pp.1-7

[156]   Kire, K., and Malhotra, N., "Software Testing using Intelligent Technique", International Journal of Computer Applications, 2014, Vol. 90, No.19

[157]   Wang, J., Zhuang, Y., Chen J., "Test Case Prioritization Technique based on Genetic Algorithm", International Conference on Internet Computing and Information Services, 2011, pp. 173–175

[158]   Ioana, L., Augustin, C., Lucia, C., "Automatic Test Data Generation for Software Path Testing using Evolutionary Algorithms", International Conference on Emerging Intelligent Data and Web Technologies, 2012, pp.1-8

[159]   Li, H., and Lam, C., "Software Test Data Generation using Ant Colony Optimization", Proceedings of ICCI, 2004 pp. 137-141

[160]   Saraph, P., Kandel, A., Last, M., "Test Case Generation and Reduction with Artificial Neural Networks", World Scientific, 2004, pp. 1-10

[161] Xie, T., "The Synergy of Human and Artificial Intelligence in Software Engineering", Proceedings of 2nd International Workshop on Realizing Artificial Intelligence Synergies in Software Engineering (RAISE), 2013

[162] Last, M. and Freidman, M., "Black-Box Testing with Info-Fuzzy Networks", 2004, Chapter 1, pp.1-25

[163] Last, M., Friendman, M. and Kandel, A., "Using Data Mining for Automated Software Testing", International Journal of Software Engineering and Knowledge Engineering, 2004, Vol. 14, pp. 369-393

[164] Shahamiri, S., and Nasir, W., "Intelligent and Automated Software Testing Methods Classification", Annual Research Seminar, 2008

[165] Ye, M., Feng, B., Zhu, L. and Lin, Y, "Neural Networks Based Automated Test Oracle for Software Testing", Springer Verlag Heidelberg, 2006, pp. 498-507

[166] Xiao, X., Xie, T., Tillmann, N., and Halleux, J., "Precise Identification of Problems for Structural Test Generation", Proceedings of ICSE, 2011, pp. 611–620

[167] Xie, T., "Cooperative Testing and Analysis: Human-tool, Tool-tool, and Human-human Co-operations to get Work Done" Proceedings of SCAM, Keynote Paper, 2002, pp. 1–3

[168] Ye, M., Feng, B., Zhu, L. and Lin, Y., "Neural Networks-based Automated Test Oracle for Software Testing", Springer Verlag, Heidelberg, 2006, pp. 498-507

[169] Mohan, V., and Jeya Mala, D., "IntelligenTester -Test Sequence Optimization Framework using Multi-Agents", Journal of Computers, 2008

[170] Lenz, A., Pozo, A., and Vergilio, S., "Linking Software Testing Results with a Machine Learning Approach", Proceedings of Engineering Applications of Artificial Intelligence, 2013, pp. 1631-1640

[171] Khoshgoftaar, T. M., Szabo, R. M. and Guasti, P. J., "Exploring the Behavior of Neural Network Software Quality Models", Software Engineering Journal, 1995, Vol. 10, pp. 89-96.

[172] Pendharkar, P., "Exhaustive and Heuristic Search Approaches for Learning a Software Defect Prediction Model", Engineering Applications of Artificial Intelligence, 2010, Vol. 23, pp. 34–40

[173] Batarseh, F., Gonzalez, A., "Predicting Failures in Agile Software Development through Data Analytics", Software Quality Journal, Springer Science, Business Media New York, 2015, pp. 49-66

[174] Afzal, W., Torkar, R., & Feldt, R., "A Systematic Review of Search based Testing for Non-functional System Properties", Journal of Information and Software Technology, 2009, Vol. 51

[175] Afzal, W., Torkar, R., & Feldt, R., "Search—based Prediction of Fault-slip-through in Large Software Projects", 2nd IEEE International Symposium on Search Based Software Engineering, 2010, pp 79–88

[176] Liu, X.F., Kane, G. and Bambroo, M., "An Intelligent Early Warning System for Software Quality Improvement and Project Management", Journal of Systems and Software, 2008, Vol. 79, No. 11, pp. 1552-1564

[177] Dweiri, F.T. and Kablan, M.M., "Using Fuzzy Decision Making for the Evaluation of the Project Management Internal Efficiency", Decision Support Systems, 2006, Vol. 42, No. 2, pp. 712-26

[178] Ryan, C., "Automatic Re-engineering of Software using Genetic Programming", Kluwer Academic Publishers, 1999

[179] Saliu, M., and Ruhe, G., "Bi-objective Release Planning for Evolving Software Systems", Proceedings of the 6th Joint Meeting of the European Software Engineering Conference and the ACM SIGSOFT International Symposium on Foundations of Software Engineering (ESEC/FSE), 2007, pp. 105–114

[180] Baker, P., Harman, M., Steinhofel, K., Skaliotis, A., "Search Based Approaches to Component Selection and Prioritization for the Next Release Problem" Proceedings of the 22nd IEEE International Conference on Software Maintenance, 2006, pp. 176–185

[181] Jiang, H., Zhang, J., Xuan, J., Re, Z., Hu, Y., 2010. A hybrid ACO algorithm for the next release problem. In: Proceedings of the 2nd International Conference on Software Engineering and Data Mining, Chengdu, pp. 166–171.

[182] Souza, J.T., Brito Maia, C.L., Ferreira, T.N., Ferreira do Carmo, R.A., Albuquerque Brasil, M. M., "An Ant Colony Optimization Approach to the Software Release Planning with Dependent Requirements", Proceedings of the 3rd International Symposium on Search Based Software Engineering (SBSE '11), 2011, pp. 142–157

[183] Alperin, L B and Kedzierski, B., "AI-based Software Maintenance", Proceedings of the 3rd IEEE Conference AI Applications, 2019

[184] Khoshgoftaar, T.M., Lanning, D.L., "A Neural Network Approach for Early Detection of Program Modules having High Risk in the Maintenance Phase", The Journal of Systems and Software, 2015, Vol. 29, pp. 85–91

[185] Teichroew, D., and Hershey, III, E., "PSL/PSA a Computer- aided Technique for Structured Documentation and Analysis of Information Processing Systems", 2nd International Conference of Software Eng., 1976, pp. 2-8

[186] Catal, C., and Guldan, S., "Product Review Management Software-based on Multiple Classifiers", Advances in Knowledge and Information Software Management, IET Software, 2017, pp. 89-92

[187] Jindal, N., Liu, B., "Review Spam Detection", 16th International Conference on World Wide Web, New York, NY, USA, 2007, pp. 1189–1190

[188] Ott, M., Cardie, C., and Hancock, J., "Estimating the Prevalence of Deception in Online Review Communities", 21st International Conference on World Wide Web, New York, NY, USA, 2012, pp. 201–210

[189] Sharma, K., Lin, K., "Review Spam Detector with Rating Consistency Check". 51st ACM Southeast Conference, New York, NY, 2013

[190] Dyer, C., "Expert Systems in Software Maintainability", Proceedings of the Annual Reliability and Maintainability Symposium, IEEE, 1984

[191] Lapa, C.M.F., Pereira, C., and De Barros, M., "A model for Preventive Maintenance Planning by Genetic Algorithms Based in Cost and Reliability", Reliability Engineering and System Safety, 2006, Vol. 91, No. 2, pp. 233-40

[192] Lethbridge T., and Singer J., "Studies of the Work Practices of Software Engineers", Advances in Software Engineering: Comprehension, Evaluation, and Evolution, Springer-Verlag, 2011, pp. 53-76.

[193] Liu, H., and Lethbridge, T., and Timothy, C., "Intelligent Search Methods for Software Maintenance", Information Systems Frontiers, ProQuest, 2002, pp. 409 - 423

[194] Mancoridis, S., Mitchell, B., Chen, Y., and Gansner, E., "Bunch: A Clustering Tool for the Recovery and Maintenance of Software System Structures", Proceedings of IEEE International Conference on Software Maintenance, 1999, pp. 50–59

[195] Russell, S., and Norvig, P., "Artificial Intelligence: A Modern Approach", Published by Prentice Hall, 3$^{rd}$ edition, 2009